\documentclass[aps,pre,preprint,superscriptaddress]{revtex4}

\usepackage{graphicx}
\usepackage{epstopdf}
\usepackage{bm}
\usepackage{amsmath}
\usepackage{mathrsfs}
\usepackage{caption}
\usepackage{subcaption}
\begin{document}


\title{Helioseismology in a bottle: Modal acoustic velocimetry}


\author{Santiago Andr\'es Triana}
\affiliation{Institute of Astronomy, KU Leuven, Belgium}
\author{Daniel S. Zimmerman}
\affiliation{IREAP, University of Maryland, USA}
\author{Henri-Claude Nataf}
\affiliation{Univ. Grenoble Alpes/CNRS/IRD, ISTerre, F-38000 Grenoble, France}
\author{Aur\'elien Thorette}
\affiliation{Univ. Grenoble Alpes/CNRS/IRD, ISTerre, F-38000 Grenoble, France}
\author{Vedran Lekic}
\affiliation{Department of Geology, University of Maryland, USA}
\author{Daniel P. Lathrop}
\affiliation{IREAP, University of Maryland, USA}

\date{\today}

\begin{abstract}
Measurement of the differential rotation of the Sun's interior is one of the great achievements of helioseismology, providing important constraints for stellar physics. The technique relies on observing and analyzing rotationally-induced splittings of {\it p}-modes in the star. Here we demonstrate the first use of the technique in a laboratory setting. We apply it in a spherical cavity with a spinning central core (spherical-Couette flow) to determine the mean azimuthal velocity of the air filling the cavity. We excite a number of acoustic resonances (analogous to {\it p}-modes in the Sun) using a speaker and record the response with an array of small microphones on the outer sphere. Many observed acoustic modes show rotationally-induced splittings, which allow us to perform an inversion to determine the air's azimuthal velocity as a function of both radius and latitude. We validate the method by comparing the velocity field obtained through inversion against the velocity profile measured with a calibrated hot film anemometer. This {\it modal acoustic velocimetry} technique has great potential for laboratory setups involving rotating fluids in axisymmetric cavities. It will be useful especially in liquid metals where direct optical methods are unsuitable and ultrasonic techniques very challenging at best.
\end{abstract}


\maketitle


\section{Introduction}
The detection of acoustic modes in the Sun led to one of the great breakthroughs in stellar physics as it allowed the determination of the internal differential rotation in the convective zone and in the upper radiative zone \cite{duvall1984internal}. Typical speeds of the turbulent flow near the Sun's surface can reach a considerable fraction of the local sound speed and thus turbulence is able to stochastically excite global acoustic resonances (called \emph p-modes) throughout the star. The eigenfrequencies of these modes are affected by rotational fluid motions in the acoustic 'cavity'. A simple way to understand this is by considering a given acoustic resonance in a non-rotating, spherically symmetric medium: its eigenfrequency cannot depend on the azimuthal wave number $m$ since the choice of the $z$ axis is completely arbitrary, but as soon as a preferred direction is introduced by rotation, the $m$ degeneracy is lifted. 

Based on a variational principle established by Chandrasekhar \cite{chandrasekhar1964general} and later generalized by Lynden-Bell \& Ostriker \cite{lynden1967stability}, it is possible to show that the eigenfrequency of an adiabatic \emph p-mode depends not only on the amount of global rotation but also on the internal differential rotation of the stellar medium \citep{gough1981new}. This is in contrast to the rotationally-induced splitting of the Earth's free oscillations, which, though routinely observed, can be successfully modeled by global rotation alone \cite{Dahlen1979rotational}. In stellar cases the \emph{forward} problem consists of the determination of the eigenfrequency shifts (splittings) of a particular eigenmode given an internal rotation law. Using rotationally-induced splittings to infer differential rotation - i.e. the \emph{inverse} problem - is inherently more difficult due to non-uniqueness. Tens of thousands of modes have been detected in the Sun allowing inversion techniques to infer with very good accuracy the internal rotation profile (see \citep{christensen2002helioseismology} for an extensive review). In stars other than the Sun a similar approach (constituting the focus of the relatively recent field of \emph{asteroseismology} \citep{aerts2010asteroseismology}) has been employed successfully, albeit with much less accuracy given the very limited number of eigenmodes observed and identified (see e.g. \citep{beck2011kepler} and \citep{deheuvels2014seismic}).

Essentially the same techniques used in helio- and asteroseismology can be used in a laboratory experiment involving a rotating fluid in an axisymmetric cavity. Conventional fluid velocimetry techniques can only give localized information in one point in space (like laser Doppler velocimetry), along a line (like ultrasound Doppler velocimetry), or on a plane (like particle image velocimetry). They all require the fluid to be seeded with neutrally buoyant tracer particles to act as scatterers (of light or ultrasound), which is expensive and not very effective particularly if the experiment needs to run for extended periods (the slightest density mismatch will make the tracers float or sink over time). We present here the first study involving acoustic modes to obtain a global measurement of a fluid's differential rotation in a spherically symmetric cavity, validated through direct measurement of the differential rotation using conventional hot film anemometry.

This study may be regarded as an experimental validation of the techniques used in helio- and asteroseismology. It provides yet another example where laboratory experiments can give valuable insight about angular momentum transport phenomena in stars, a fundamental factor in their evolution that has gained attention recently (see e.g. \citep{ruediger2014angular} and \citep{mathis2014impact}).   

We will describe the experimental setup and measurements, followed by a description of the data analysis including a 2D inversion and validation. We conclude by discussing the potential of the technique as a general diagnostic tool for rotating fluids in axisymmetric cavities.

\section{The experiment}
\subsection{Setup}
\begin{figure}[t!]
\includegraphics[width=0.7\linewidth]{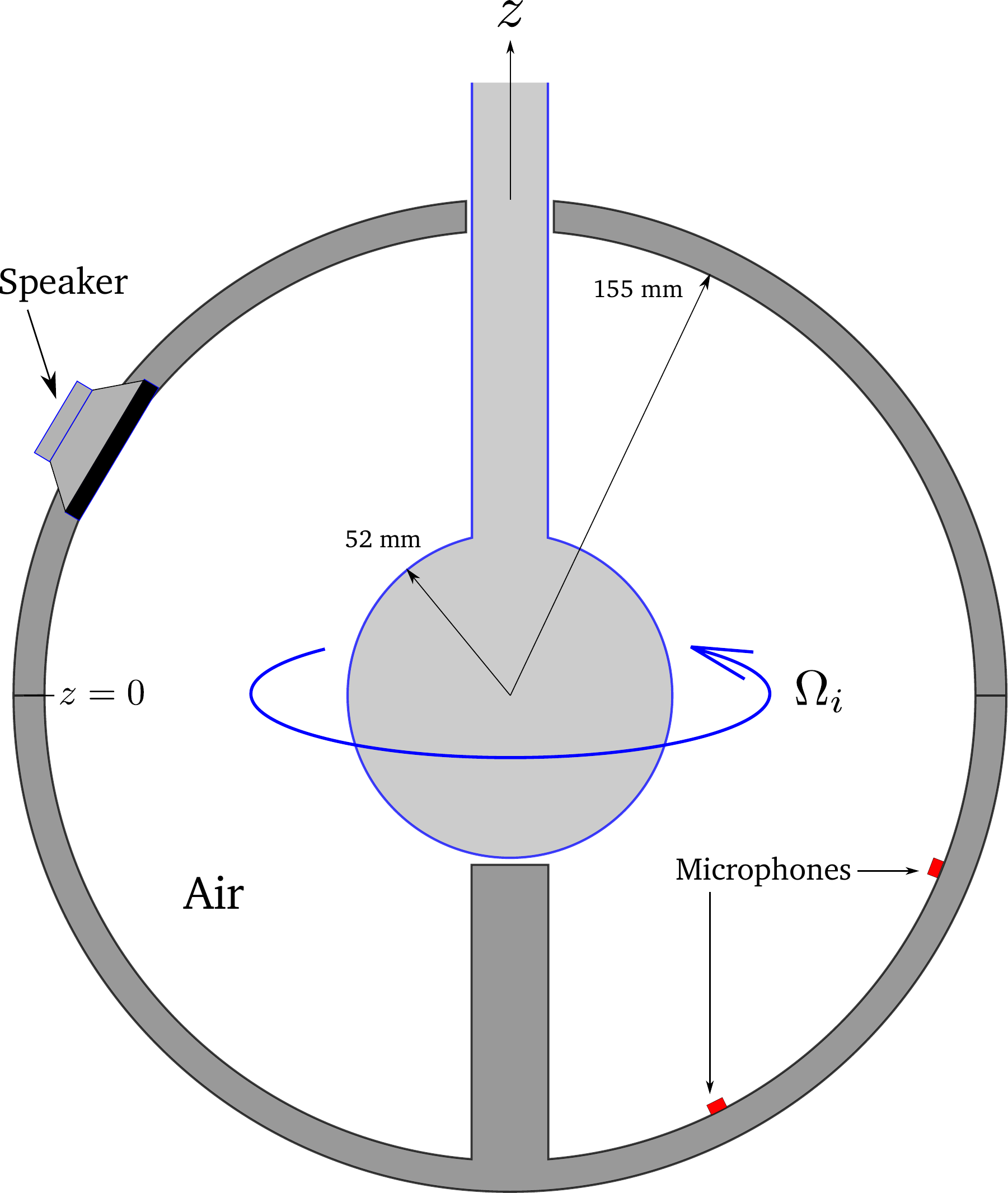}
\caption{\label{fig:30cmdiag} Meridional cross section of the experimental setup. The inner sphere can spin driven by an electric motor (not shown). The speaker provides the acoustic excitation and the response is recorded with the microphone array. Below the inner sphere there is a cylinder attached rigidly to the cavity (matching the inner sphere's shaft dimensions) with the sole purpose of maintaining equatorial symmetry as much as possible.}
\end{figure}

The experiment was performed at the University of Maryland's GeoDynamo Lab. The device, reused from a former liquid sodium experiment, consists of a spherical shell cavity between an outer stainless-steel shell (inner radius $r_o=155\pm3$ mm), stationary in the lab frame, and an inner copper sphere (radius $r_i=52\pm1$ mm), concentric with the outer and supported by a $25$ mm diameter shaft (see Fig. \ref{fig:30cmdiag}). The inner sphere can spin up to 36 revolutions per second driven by an electric motor connected to the shaft. The outer sphere has a 50 mm wide speaker fitted through an opening at $60^\circ$ colatitude. An array of six Panasonic WM-61A electret microphones (their active component is a dielectric material with a permanently embedded static electric charge) on the lower hemisphere records the pressure response of the air filling the cavity when driven acoustically by the speaker. Four of the microphones are located at about $120^\circ$ colatitude and separated $22.5^\circ$ evenly in the azimuthal direction. The remaining two are at $160^\circ$ colatitude, are separated $67.5^\circ$ azimuthally to match meridionally with the end microphones located at $120^\circ$ colatitude above (see Fig. \ref{fig:30cmdiag}). Placement of the source speaker and microphones determines which modes can be excited and observed. It is important to identify the optimal microphone/speaker placement for a particular application; in this paper, we do not attempt to devise a scheme for doing so.

During a typical experimental run, the inner sphere is brought to a constant spinning rate $f_i = \Omega_i/2\pi$, with $\pm 0.5$ Hz accuracy, and maintained for a 10-minute interval. During the last six minutes of that interval, the speaker is driven by a superposition of sinusoidal signals whose frequencies change slowly and linearly in time (i.e. chirps). Each chirp sweeps a frequency window of about 200 Hz, targeting a particular eigenmode family or families. Overall they cover frequencies from 400 Hz up to 6 kHz. The acoustic response of the cavity as registered by the microphone array is recorded by means of an audio interface (M-Audio Fast Track Pro, 48 kHz sampling rate) connected to a computer.

\subsection{Measurements}
\begin{figure*}[t]
	\centering
	\begin{subfigure}[b]{\linewidth}
		\includegraphics[width=\linewidth]{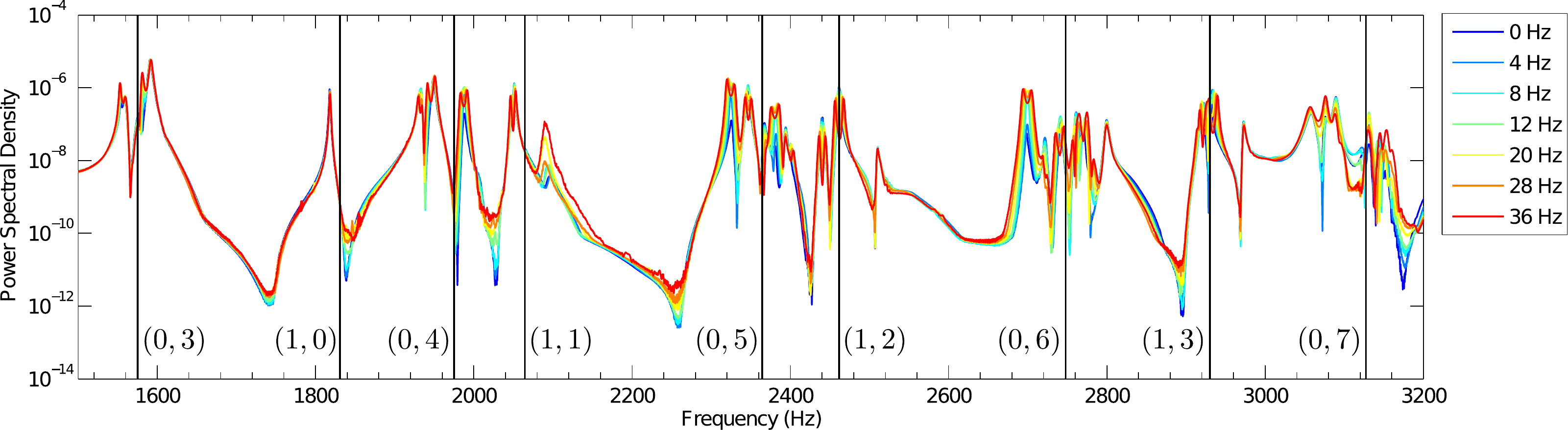}
		\caption{\label{fig:spect1}Acoustic response of the spherical shell cavity as recorded through one of the microphones near the equator. The spectrum shown here is the average from four repetitions of a single continuous chirp. Color indicates the rotational speed of the inner sphere and the vertical lines mark the theoretical eigenfrequencies labeled $(n,l)$.}
	\end{subfigure}\vspace{5mm}
	\begin{subfigure}[b]{\linewidth}
		\includegraphics[width=\linewidth]{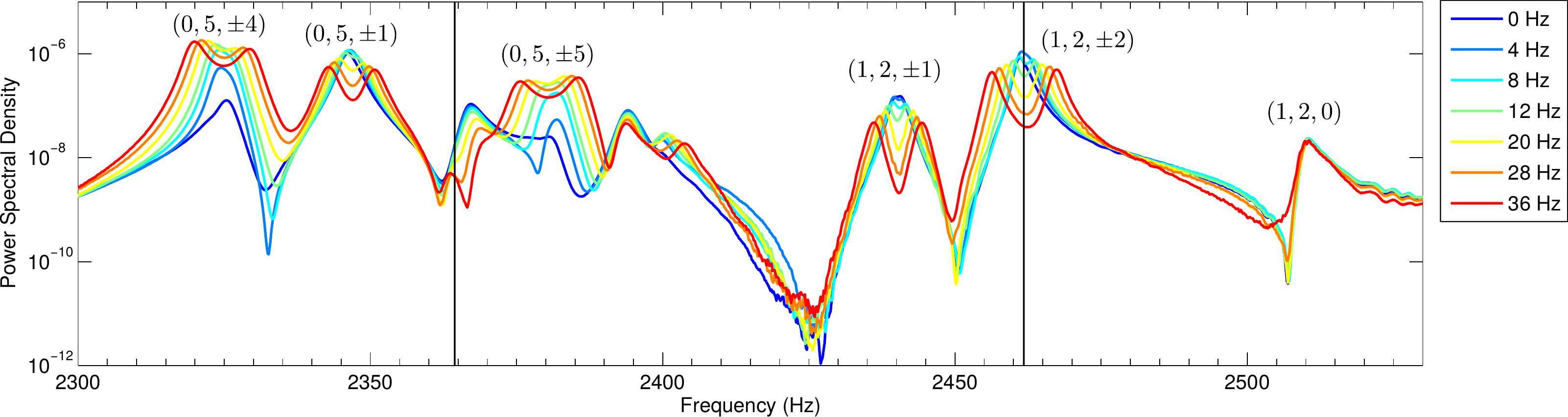}
		\caption{\label{fig:spect2}Detail from Fig.\ref{fig:spect1} showing the frequency splitting induced by rotation of the inner sphere. The resonance family to the left corresponds to $(n=0,l=5)$ modes where only pairs with $|m|=1,4,5$ could be identified. The family on the right corresponds to $(n=1,l=2)$.}
	\end{subfigure}
        \caption{\label{fig:spect}Acoustic spectra}
\label{fig:spectra}
\end{figure*}

We recorded pressure signals with the inner sphere stationary and up to 36 Hz increasing in 4 Hz intervals, while the outer spherical container remained at rest. Since we were interested in the resonant frequencies and not in the modes' amplitudes, a proper calibration of the microphone array was not needed. The acoustic power forced in the cavity by the chirp was therefore not measured but we did compare the power of the recorded pressure with and without the chirp resulting typically in a 20 dB ratio, i.e. the sound levels from the flow itself (and the rest of the apparatus) were 20 dB lower than the acoustic forcing by the speaker. Figure \ref{fig:spectra} shows the typical spectra (obtained by averaging the spectra from four repetitions of a single-continuous chirp) computed from one of the microphones. Performing more repetitions would improve the signal to noise ratio of the spectra, but as we explain below, this would require more measurement time during which the temperature might drift, affecting slightly the eigenfrequencies. In principle all eigenmodes up to 6 kHz in frequency are excited, but given the fixed location of the speaker, which might lie on a nodal line for some modes, the corresponding amplitudes would be much smaller compared to other modes.

To obtain the splittings we measured the separation in frequency of the splitting pairs peaks' maximum (after proper identification) from each microphone's signal and averaged across microphones. We estimated the error from the corresponding standard deviation or the resolution of the power spectrum, whichever the greater. Splittings from all modes appear to be linear with the rotation rate of the inner sphere, see Fig. \ref{fig:splits}.

\begin{figure}
\includegraphics[width=0.8\linewidth]{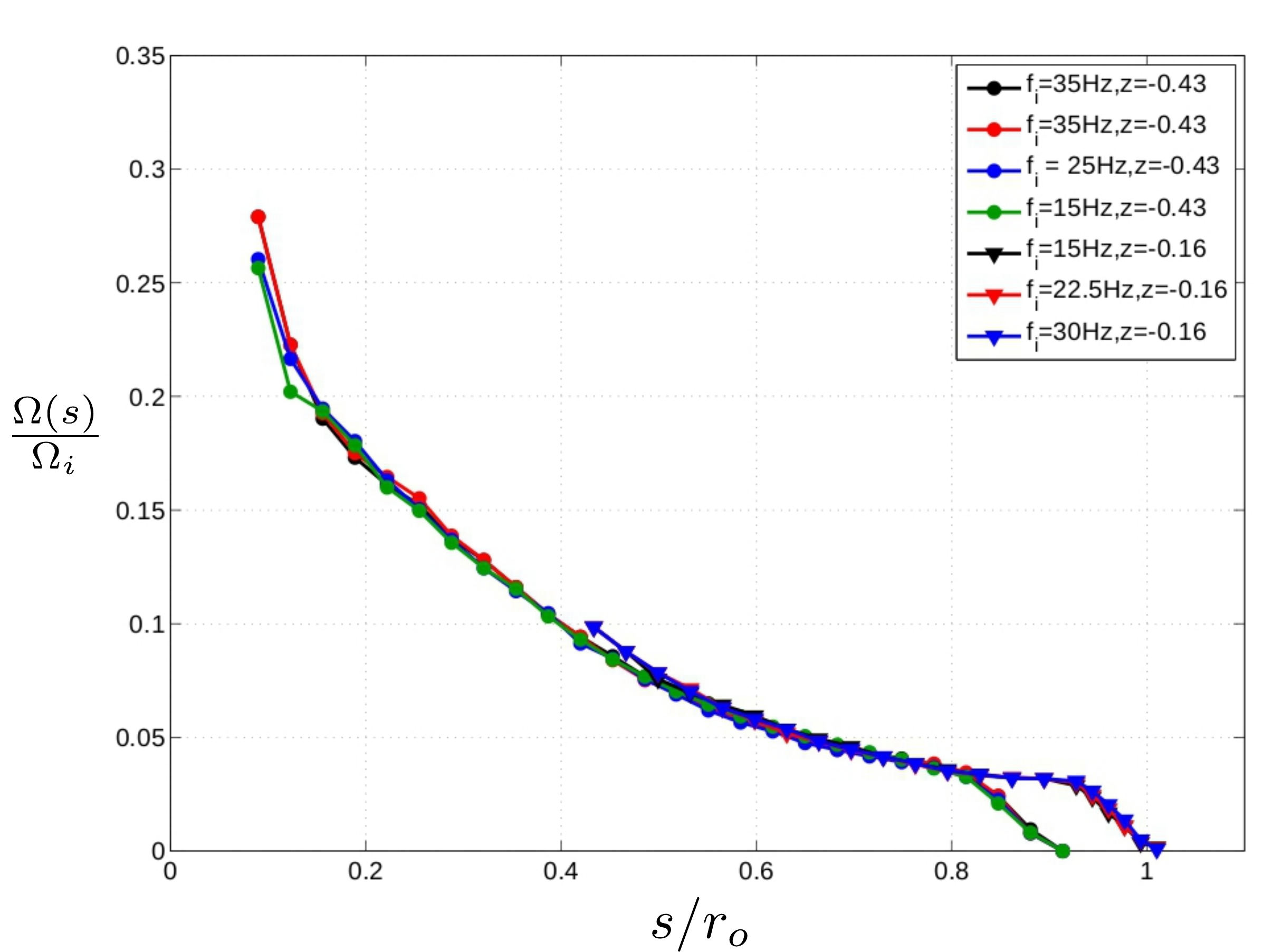}
\caption{\label{fig:30cmprof} The azimuthal flow velocity $U_\phi$ from measurements with a hot film anemometer, shown here normalized with $s \Omega_i$, and plotted as a function of the cylindrical radius $s$ for different values of $f_i = \Omega_i/2\pi$ measured at two different distances from the equator ($z=0.16$ and $z=0.43$).}
\end{figure}

\begin{figure*}
\includegraphics[width=0.9\linewidth]{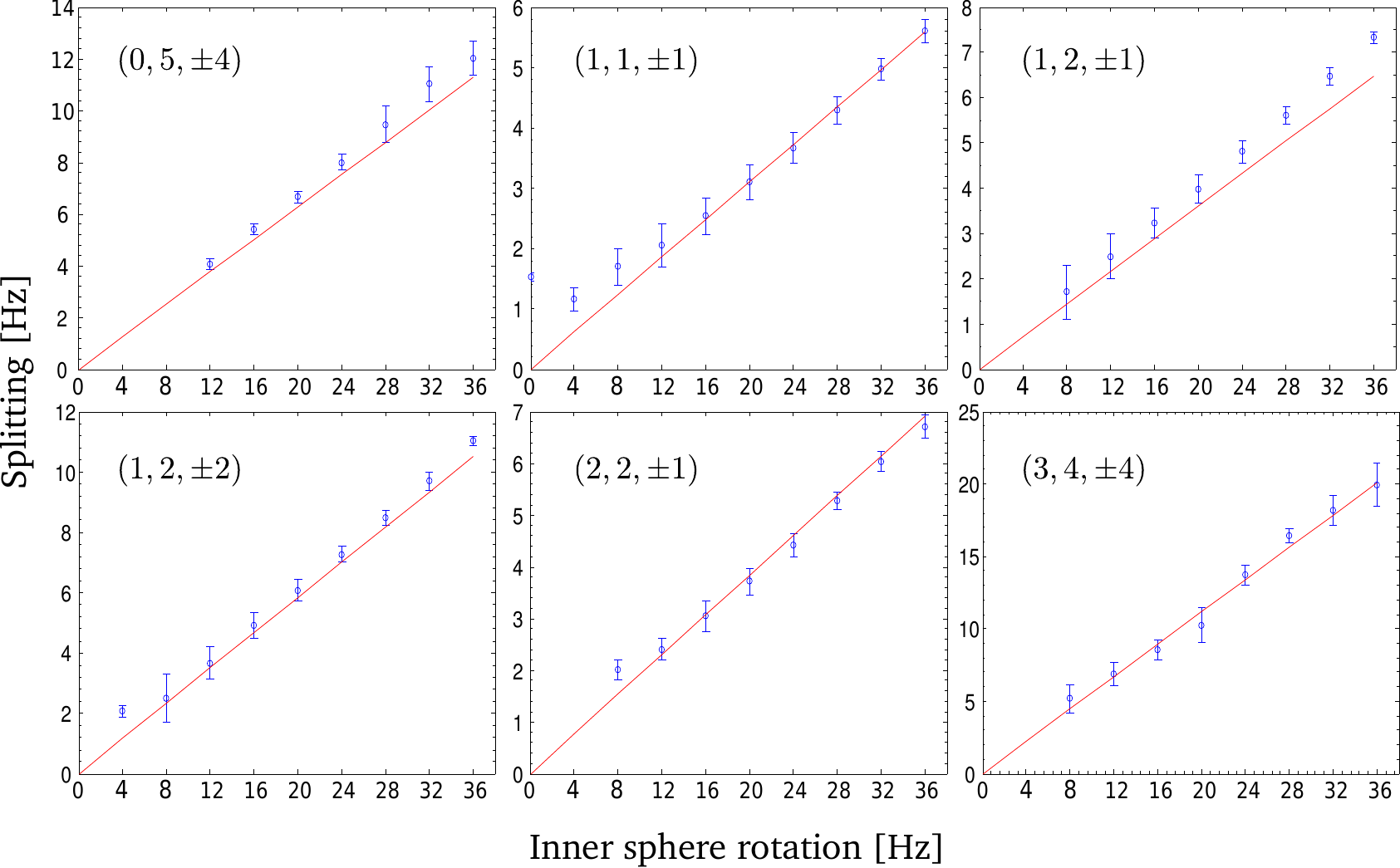}
\caption{\label{fig:splits} Sample of six mode splitting measurements as functions of the inner sphere rotation rate. The red straight line is the theoretical splitting from Eq. \ref{eq:split} based on the measured azimuthal velocity profile shown in Fig. \ref{fig:30cmprof}, assuming $z$-invariance.}
\end{figure*}

We also performed measurements along two cylindrical radii of the flow's azimuthal velocity using a calibrated hot film anemometer  (TSI model 1650, with a 4.8 mm diameter tip). Experimental parameters were identical except perhaps from the small obstruction of the flow from the hot film probe itself. We measured azimuthal velocity profiles for all inner sphere rotation rates along cylindrical radii at two different heights, $z=0.16$ and $z=0.43$ (using $r_o=155$ mm as the unit length). The resulting velocity profiles are shown in Fig. \ref{fig:30cmprof}. The profiles are largely invariant along the $z$ axis, likely as a result of a geostrophic balance. The velocity profiles appear to scale linearly with the rotation rate of the inner sphere, which reflects on the linear dependence of the splittings. A discussion of the physics underlying the velocity profiles is beyond the scope of this paper, which concerns itself with ways of measuring them.

\section{\label{sec:mi}Mode identification and forward problem}
Having excited and recorded the resonances, we must identify them with particular acoustic eigenmodes. To do this, we use a non-rotating model as reference. We approximate the cavity as a spherical shell (ignoring the shaft supporting the inner sphere and the cavity's deviations from sphericity) and calculate semi-analytically the pressure eigenmodes, a straightforward task that involves solving the Helmholtz equation subject to the condition that the radial derivative of the pressure $\partial P/\partial r$ is zero at both $r=r_i$ and $r=r_o$ (see Appendix \ref{ap:acmo}). A given eigenmode can be identified by three indices $(n,l,m)$ where $n$ is the radial order and $l,m$ are the indices of the spherical harmonics $Y_l^m(\theta,\phi)$. These theoretical eigenfrequencies are plotted as vertical lines in Fig. \ref{fig:spectra}.

In a perfectly spherically-symmetric model, the $2l+1$ modes with the same $(n,l)$ numbers oscillate at the same frequency, i.e. they are degenerate. The presence of the shaft and small deviations from sphericity in the real setup (the outer sphere is slightly prolate along the $z$ axis) introduces a preferred axis even when the inner sphere is at rest. This causes modes with the same $(n,l)$ numbers but different \emph{absolute} values $|m|$ to be shifted differently, i.e. each $(n,l)$ family splits into an $l+1$ multiplet. This can be seen in the non-rotating (0 Hz) spectra in Fig. \ref{fig:spectra}.

When the inner sphere spins, each $(n,l,\pm m)$ mode pair splits due to rotation as expected, resulting finally in a $2l+1$ multiplet for each $(n,l)$ mode family. The theoretical eigenfrequencies can be associated with the observed mode families, at least at low frequencies where the mode frequency spacings are wide enough. Determination of the azimuthal wave number $m$ of a particular resonance was achieved by computing the time-delayed cross-correlation of signals (band-passed around the resonance) recorded by microphones at different azimuthal locations. Through this process we were able to identify unambiguously a total of $M=26$ acoustic modes undergoing rotationally-induced splitting.

During the experimental runs the temperature of the air in the cavity increased slightly (about $2^\circ$C above the initial temperature) thus increasing the sound speed. This temperature increase is due to heat produced by the air's friction against the spinning inner sphere and accumulated over time. The effect was easily quantified by the slight frequency drift (close to 0.3\%) of $m=0$ modes (they are completely unaffected by the inner sphere rotation, apart from the temperature increase). We monitored the mode $(n=1,l=0,m=0)$ at all times during data gathering and used it as a reference to compare spectra taken with drifting temperatures. Since the wavenumber $k=2\pi f/c$ is fixed and knowing the functional dependence of the sound speed $c$ with temperature, it is possible to estimate with high precision the temperature of the air in the cavity ($\pm 0.1^\circ$C). Other higher frequency $m=0$ modes besides $(1,0,0)$ can be used in principle. The higher the frequency of the reference mode, the smaller the error on the calculated temperature. The accuracy of this temperature estimation is of course tied to the general accuracy of the model used to compute the eigenfrequencies.

The measured rotational profiles $\Omega(s,z=0.16)$ and $\Omega(s,z=0.43)$, when scaled by the inner sphere rotation rate (Fig. \ref{fig:30cmprof}), suggest a $z$-independent profile. We can use them to calculate the \emph{expected} rotational splitting for the modes we identified and compare with the measured splittings. For this task it is necessary to calculate the \emph{kernels} $K_{nlm}$ from the theoretical pressure eigenfunctions (see Appendix \ref{ap:kern} for details).
The (angular) frequency shift of a particular mode $(n,l,m)$ is given by (if $\Omega/2\pi$ is small compared to the eigenfrequency)
\begin{equation}
\Delta_{nlm}= m \int_{r_i}^{r_o} \int_0^\pi K_{nlm}(r,\theta)\, \Omega(r,\theta)\, r {\rm d} r {\rm d} \theta,
\label{eq:split}
\end{equation}
which can be computed using our measured $\Omega$ (assuming it independent of $z$). For a meaningful comparison with the splitting measurements, we compute first a linear fit (slope and intercept) and subtract the intercept from each measurement. This way we eliminate any splitting bias that might be present even with the inner sphere at rest, i.e. we consider only the splitting due to an increase in the rotation rate of the inner sphere. The result of this exercise is shown in Fig. \ref{fig:splits}, where it is clear that the observed (slope) and predicted splittings agree reasonably well. Having established the consistency between the forward problem and the measured profiles, tackling the inverse problem becomes feasible.

\section{Inversion}
\subsection{Tikhonov regularization}
We can now attempt an inversion to obtain an approximated 2D rotational profile $\bar \Omega(r,\theta)$ using the measured mode splittings $\Delta_i$ ($i=1,\ldots, 26$). One possible approach relies on a Tikhonov regularization method commonly used in helioseismology. This method estimates $\bar \Omega(r,\theta)$ by minimizing the quantity $\mathscr{T}$ defined as
\begin{equation}
\mathscr{T}=\sum_{i=1}^{26} \frac{\left(\Delta_i - \bar \Delta_i \right)^2}{\epsilon_i^2}+
\mu_r \int_{r,\theta}\left(\frac{\partial^2 \bar \Omega}{\partial r^2}\right)^2 \mathrm{d}r\, \mathrm{d}\theta +
\mu_\theta \int_{r,\theta}\left(\frac{\partial^2 \bar \Omega}{\partial \theta^2}\right)^2 \mathrm{d}r\, \mathrm{d}\theta,
\label{eq:Tikh0}
\end{equation}
where $\bar \Delta_i$ are the predicted splittings, $\epsilon_i$ is the measurement error and $\mu_r,\mu_\theta$ are parameters controlling the smoothing on the second-derivatives. To begin, we consider a grid of $N_r+1$ points in the radial direction and $N_\theta+1$ points in the $\hat \theta$ direction so that $\bar \Omega$ discretizes into $N_r\times N_\theta$ values to be determined, translating the minimization problem into a matrix inversion problem, see Appendix \ref{ap:tr} for details.

\begin{figure}
\includegraphics[width=\linewidth]{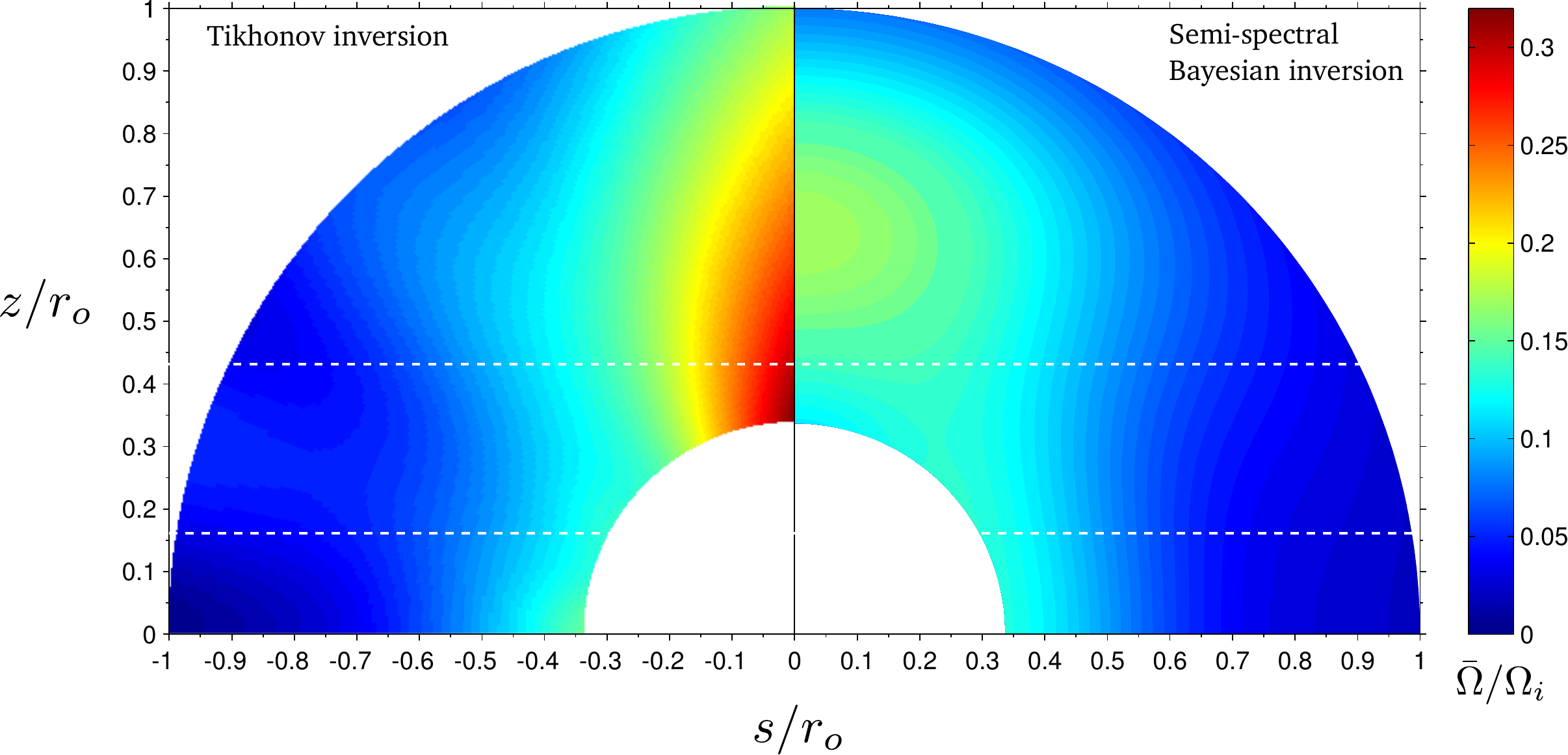}
\caption{\label{fig:oi} On the left quadrant the angular velocity $\bar \Omega$ as computed through a 2D Tikhonov inversion technique with trade-off parameters $\mu_r=10^{-3}$ and $\mu_\theta=2\times 10^{-5}$. There are $Nr=100$ radial points and $N_\theta=180$ points in the $\hat \theta$ direction. The quadrant on the right shows the results from a semi-spectral Bayesian inversion. The two horizontal dashed lines indicate the location of the $U_\phi$ velocity profile measurements from anemometry shown in Figs. \ref{fig:30cmprof} and \ref{fig:comparison}. The two inversion techniques give very similar results except in the regions close to the rotation axis where the splitting modes provide no information.}
\end{figure}

An inversion using trade-off parameters $\mu_r=10^{-3}$, $\mu_\theta=2\times 10^{-5}$ (see below on how to choose these values) and $N_r=100$, $N_\theta=180$ grid points, as calculated from Eq. \ref{eq:TO}, is shown in the left quadrant of Fig. \ref{fig:oi}.   

\begin{figure}
	\centering
	\begin{subfigure}[b]{0.9\linewidth}
		\includegraphics[width=0.9\linewidth]{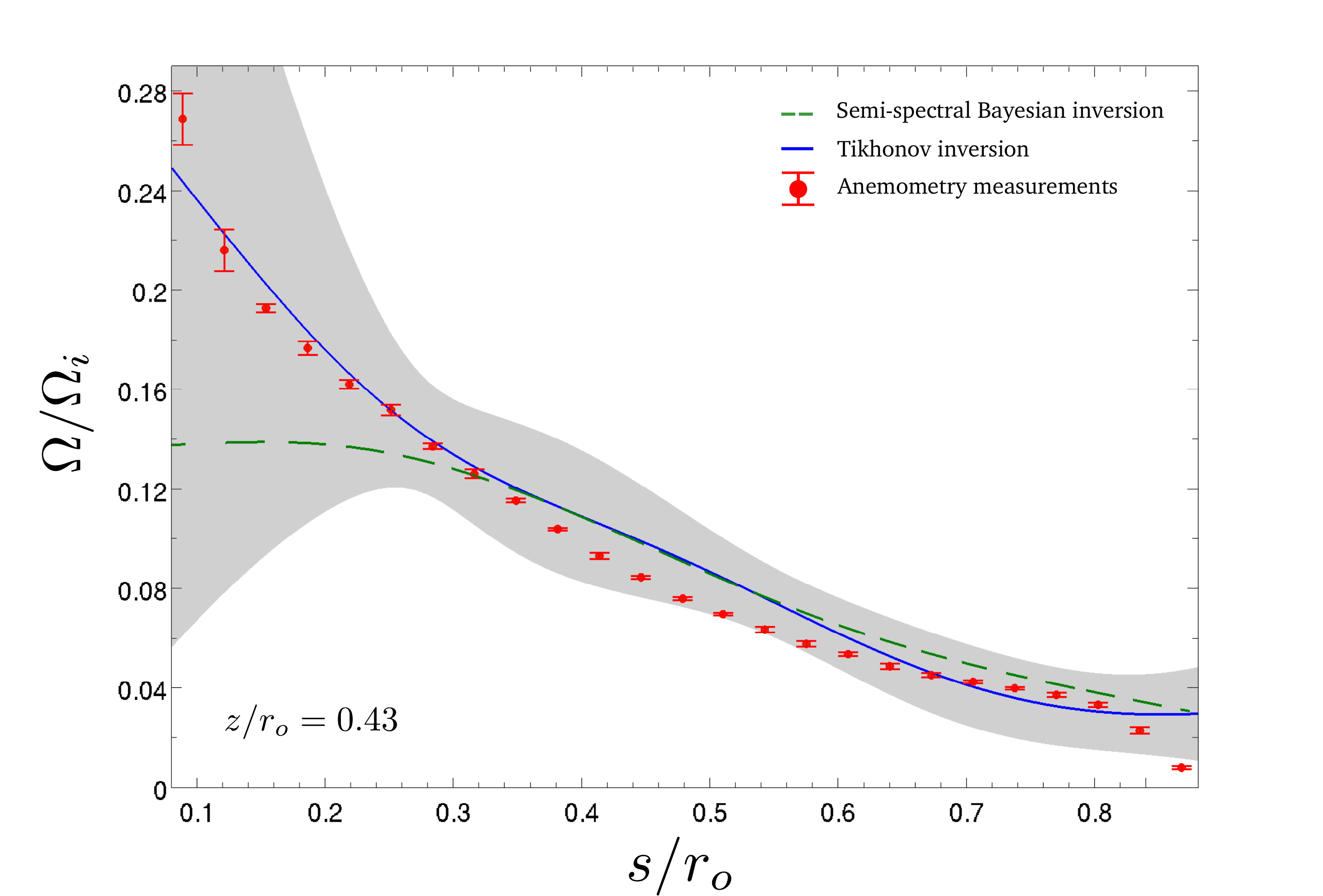}
		\caption{\label{fig:o16} Scaled angular velocity at $z/r_o=0.43$}
	\end{subfigure}
	\begin{subfigure}[b]{0.9\linewidth}
		\includegraphics[width=0.9\linewidth]{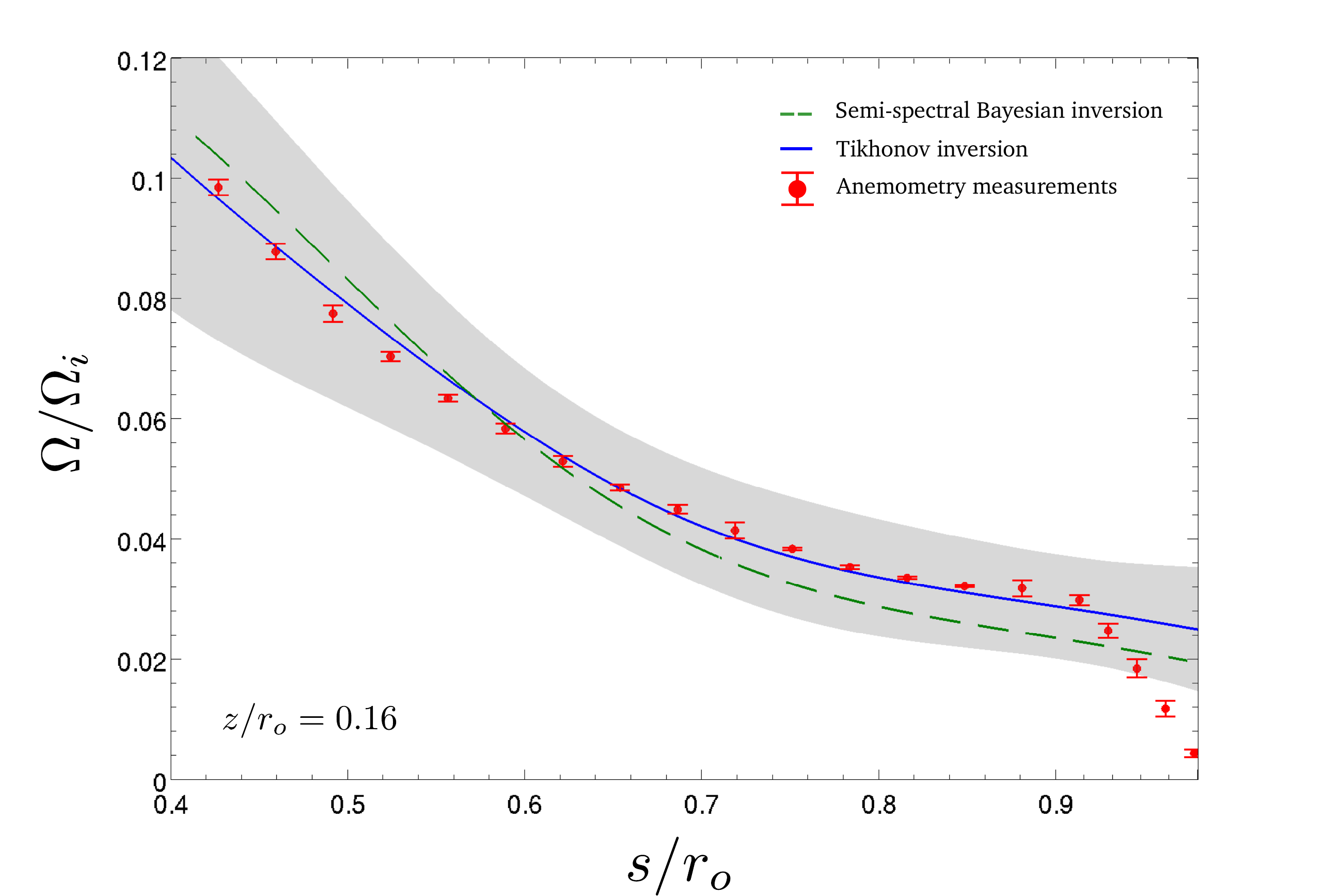}
		\caption{\label{fig:o43} Scaled angular velocity at $z/r_o=0.16$}
	\end{subfigure}
        \caption{\label{fig:o}Comparison between the $\Omega(s)$ profiles measured experimentally at the locations shown in Fig. \ref{fig:oi} and $\bar \Omega$ obtained through inversion. Gray shaded area indicates the $1\sigma$ error on the Tikhonov inversion.}
\label{fig:comparison}
\end{figure}

In Fig. \ref{fig:comparison} we compare the inversion result with the $\Omega$ profile measured experimentally at two different heights. To make a proper comparison we interpolate $\bar \Omega$ at the different cylindrical radius $s$ where the measurements were performed. The {\it rms} of the difference at $z=0.16$ is just $6\times 10^{-3}$ and $10^{-2}$ at $z=0.43$. 

We see that the technique does a reasonably good job at approximating the true angular velocity, especially considering the reduced number of splittings $(M=26)$ that we were able to identify and measure.

\begin{figure}
	\includegraphics[width=0.8\linewidth]{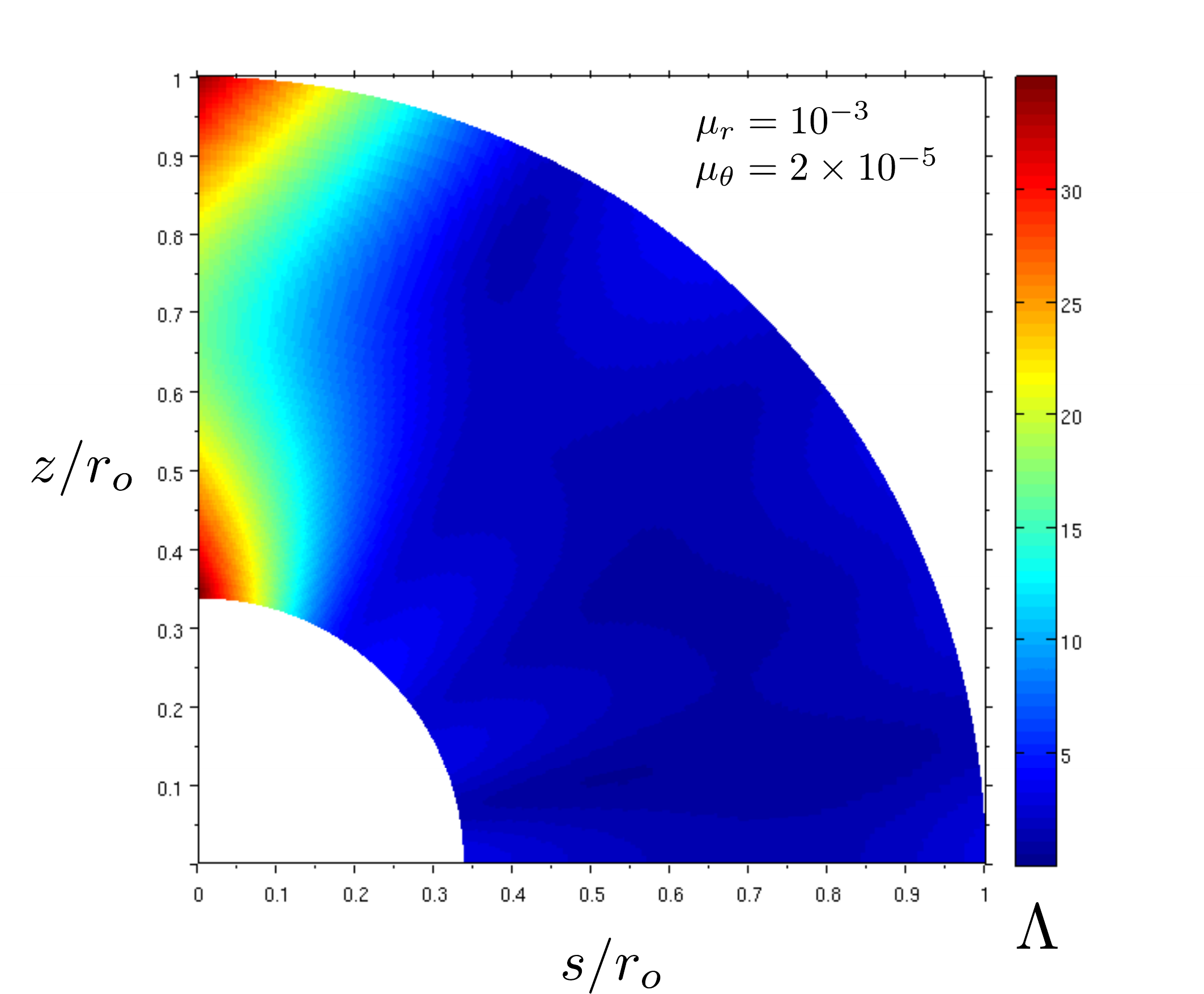}
	\caption{\label{fig:ermag}Error magnification $\Lambda(s,z)$ corresponding to a Tikhonov inversion with trade-off parameters $\mu_r=10^{-3}$ and $\mu_\theta=2\times 10^{-5}$. Regions close to the rotation axis are reconstructed poorly by the inversion.}
\end{figure}

The spatial resolution and errors involved in the inversion are usually discussed in terms of the averaging kernels $\mathcal{K}(r',\theta', r,\theta)$ and the error magnification $\Lambda(r',\theta')$, respectively. Ideally $\mathcal{K}$ is a sharply peaked function at $(r=r',\theta=\theta')$ and close to zero everywhere else, its width is a measure of the spatial resolution (see Eqs. \ref{eq:inv_coeff} and \ref{eq:avker}). The error magnification is a measure of the uncertainty on $\bar \Omega$ (see eq. \ref{eq:ermag}). Let us examine $\mathcal{K}$ and $\Lambda$ for this particular inversion. The error $\Lambda$ at a given $(r',\theta')$ is obtained by the $\sqrt{M}\times$ {\it rms} of the corresponding row of $\mathbf{T}^{-1}\mathbf{Q}^\top$ (each row corresponds to a grid interval, each column to a mode). The result is shown in Fig. \ref{fig:ermag}. It is worth noting that a resolution matrix analysis, used often in seismic tomography, provides analogous information as $\mathcal{K}$, and suffers from similar limitations, such as the assumption of accurate forward modeling.

\begin{figure}
\includegraphics[width=\linewidth]{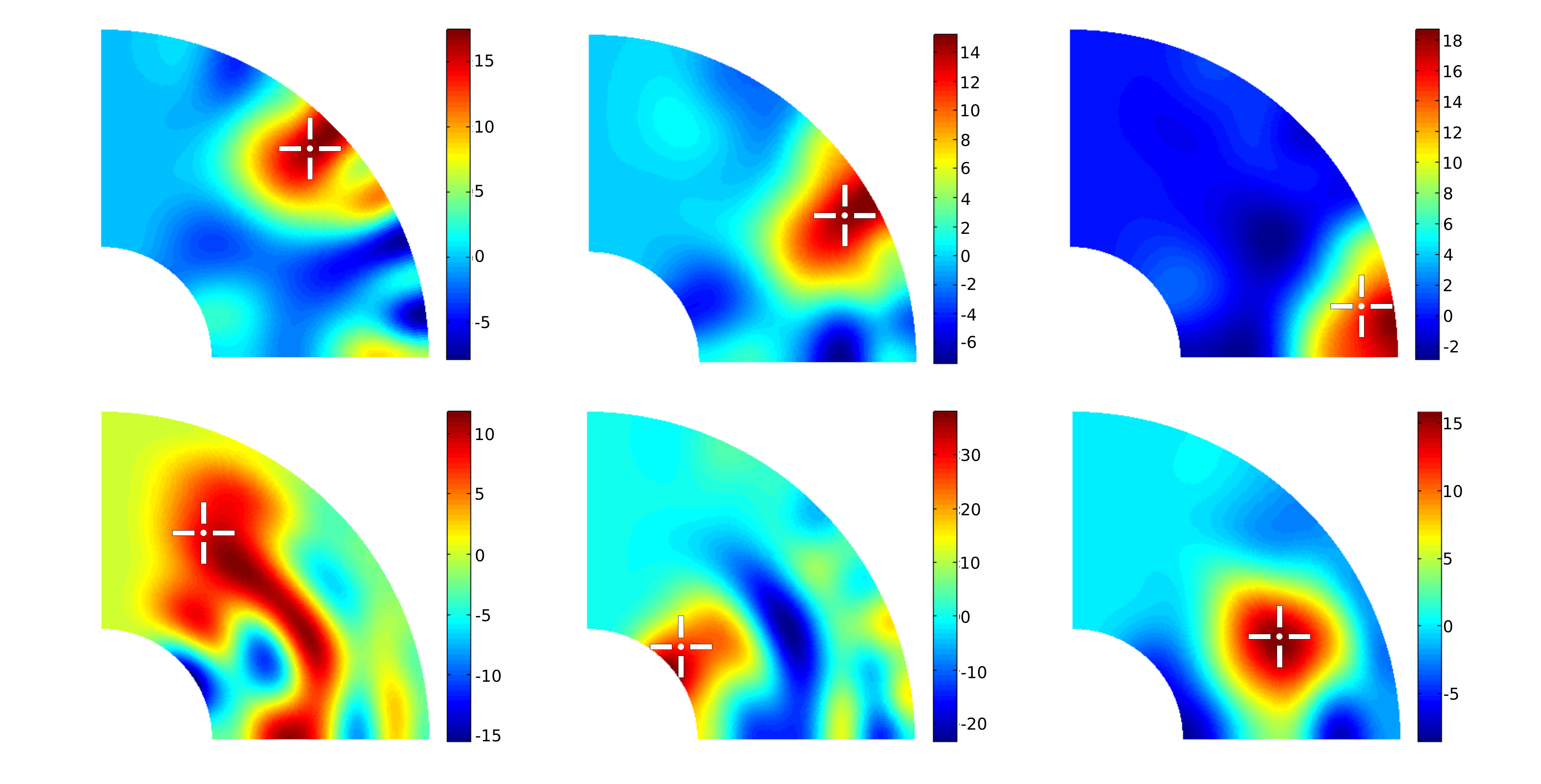}
\caption{Averaging kernels for various $(r', \theta')$ points indicated by the white cross marks. Figure at bottom, right corresponds to the same  $(r', \theta')$ point as Figs \ref{fig:trade_ermag} through \ref{fig:trade_aw}. Trade-off parameters are $\mu_r=10^{-3}$ and $\mu_\theta=2\times 10^{-5}$. Color scale is dimensionless.} 
\label{fig:avker}
\end{figure}

Regions close to the rotation axis have significant error magnification (since angular velocities near the axis have little effect on the modes), while other regions show values of $\Lambda$ close to or less than one. The averaging kernels $\mathcal{K}$ for a sample of six $(r',\theta')$ target points are shown on Fig. \ref{fig:avker}. The kernels are generally centered close to their associated $(r',\theta')$ target point, with sizable width both in the $\hat r$ and in the $\hat \theta$ direction. When the measured mode splittings do not contain sufficient information for a reliable inference of flow velocity at a target location, the averaging kernels appear peaked in several places away from the target location (e.g. Fig. \ref{fig:avker}, bottom left).

\begin{figure*}
	\centering
	\begin{subfigure}[b]{0.45\linewidth}
		\includegraphics[width=\linewidth]{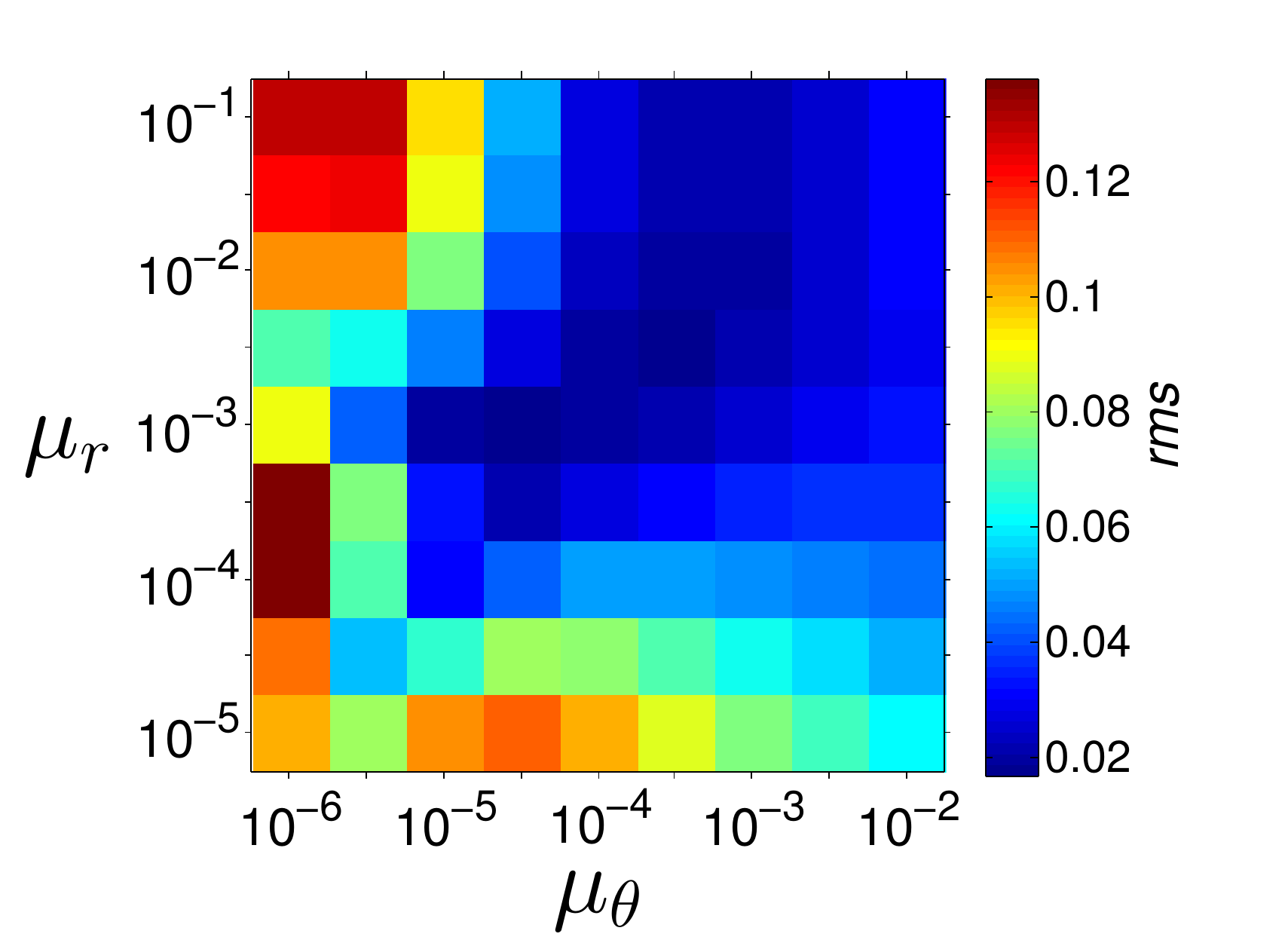}
		\caption{rms}
		\label{fig:trade_rms}
	\end{subfigure}\quad
	\begin{subfigure}[b]{0.45\linewidth}
		\includegraphics[width=\linewidth]{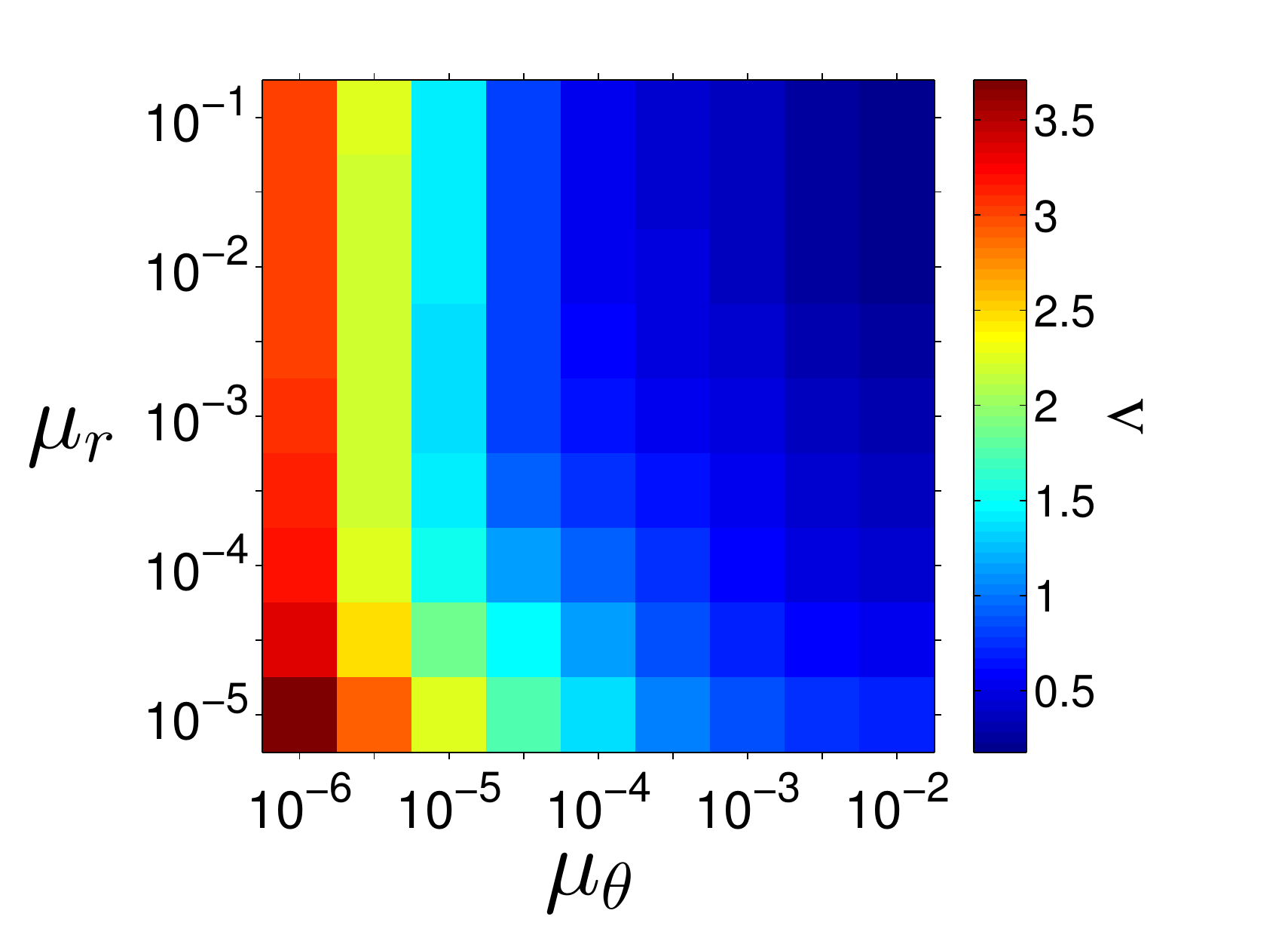}
		\caption{Error magnification $\Lambda$}
		\label{fig:trade_ermag}
	\end{subfigure}

	\begin{subfigure}[b]{0.45\linewidth}
		\includegraphics[width=\linewidth]{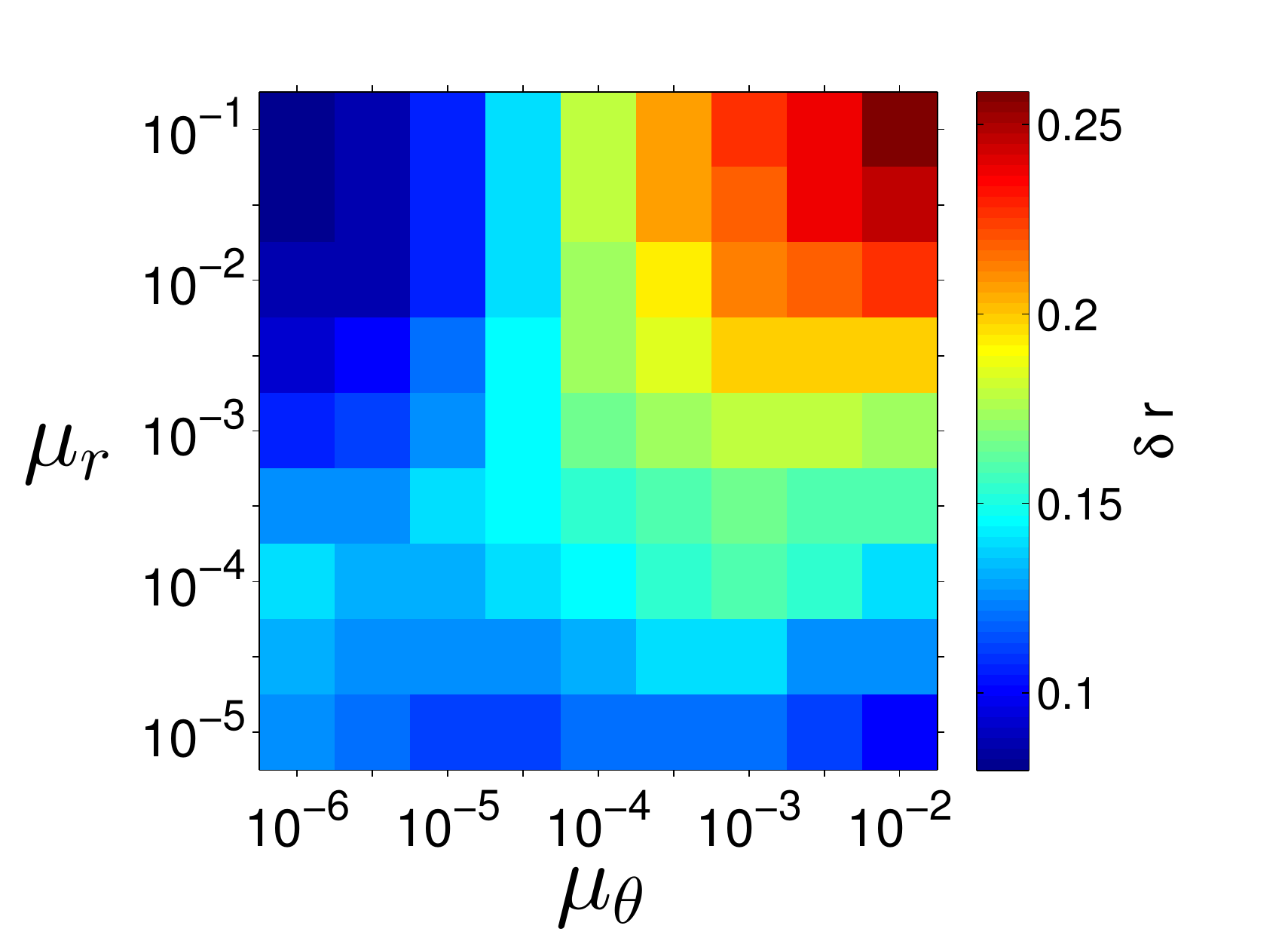}
		\caption{Radial width}
		\label{fig:trade_rw}
	\end{subfigure}\quad
	\begin{subfigure}[b]{0.45\linewidth}
		\includegraphics[width=\linewidth]{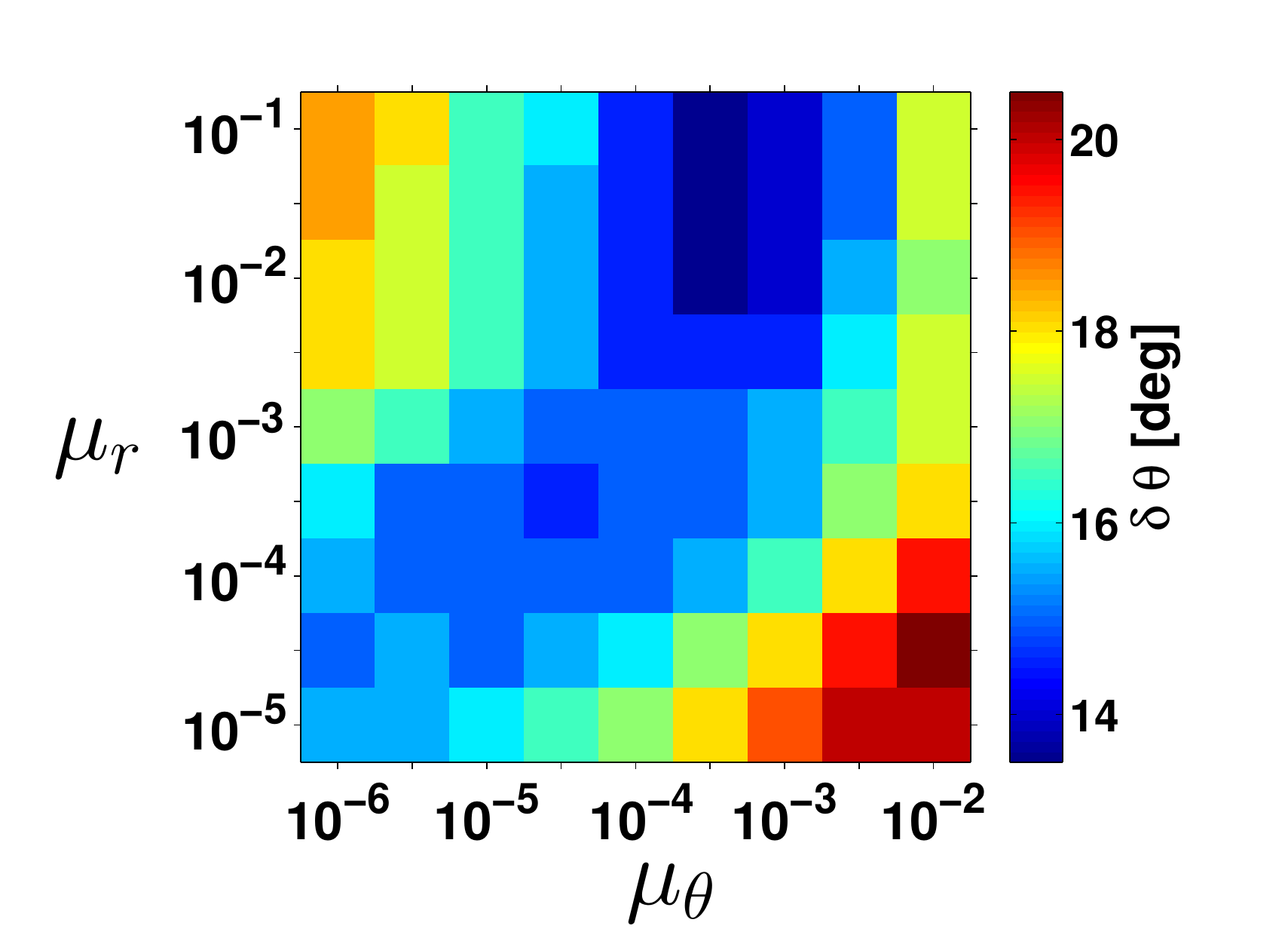}
		\caption{Angular width}
		\label{fig:trade_aw}
	\end{subfigure}
        \caption{Wide range variation of the trade-off parameters $\mu_r$ and $\mu_\theta$. The overall deviation from the hot-film anemometry profiles is shown in \ref{fig:trade_rms}. The error magnification at $r'=0.7$, $\theta'=64^\circ$ is shown in Fig. \ref{fig:trade_ermag}, while Figs. \ref{fig:trade_rw} and \ref{fig:trade_aw} show the radial and angular width of the averaging kernels at that particular $(r',\theta')$.}\label{fig:trade}
\end{figure*}

The uncertainty on the inferred $\bar \Omega$ is provided by Eq. \ref{eq:acc}, and trades off with the smoothing that is imposed; larger smoothing results in reduced lateral resolution but increased confidence on $\bar \Omega$, lower smoothing results in improved lateral resolution, but decreased confidence on $\bar \Omega$. To see this, let us select the averaging kernel $\mathcal{K}$ corresponding to the target point $(r'=0.7,\theta'=64^\circ)$ (Fig. \ref{fig:avker}, bottom right) to study how the radial and angular widths (defined by Eq. \ref{eq:width}) change as we vary the trade-off parameters over a wide range. This is shown in Fig. \ref{fig:trade}, along with the error magnification $\Lambda(r'=0.7,\theta'=64^\circ)$ and the sum of the {\it rms} deviation of $\bar \Omega$ from the anemometry measurements.

The trade-off parameters are usually chosen so that there is an adequate compromise between resolution and error. An error magnification $\Lambda$ close to one is often a good choice representing such compromise (Fig. \ref{fig:trade_ermag}). If we desire an error magnification close to one at the chosen $(r',\theta')$, while achieving a balance between radial and angular resolution, we see from Figs. \ref{fig:trade_rw} and \ref{fig:trade_aw} that the optimal choices for $\mu_r$ and $\mu_\theta$ are in the vicinity where the {\it rms} deviations are the smallest (Fig. \ref{fig:trade_rms}). 

\subsection{Semi-spectral Bayesian inversion}

The Tikhonov regularization method has the advantage of being easily generalized to a non-spherical geometry. However, for our current problem, a more simple and lighter method makes use of the fact that the spherical harmonics form a natural basis for both the acoustic modes and the flow field.
More specifically, the azimuthal velocity field $U_\phi(r,\theta)= \Omega(r,\theta) \, r \sin \theta$ can be written as \citep{nataf13magnetic}:

\begin{equation}
U_\phi(r,\theta) = \sum_{odd \; l''=1}^{l_{max}} U_{l''}(r) P_{l''}^1(\cos \theta),
\label{eq:U_phi}
\end{equation}
where $P_{l}^1$ is the associated Legendre polynomial of degree $l$ and order $1$.
Injecting this expression and the similar expansion (\ref{eq:Legendre_K}) for the rotational kernels into equation \ref{eq:split}, we obtain the splitting $\Delta_{nlm}$ as:


\begin{equation}
\Delta_{nlm} = m \int_{r_i}^{r_o}dr \int_0^{\pi}  \sum_{odd \, l'=1}^{2l+1} \sum_{odd \, l''=1}^{l''_{max}} K_{l'}(r) U_{l''}(r)\frac{P_{l'}^1(\cos\theta) P_{l''}^1(\cos\theta)}{\sin\theta} d\theta,
\label{eq:integration_1}
\end{equation}
which simplifies into:
\begin{equation}
\Delta_{nlm} = m \sum_{odd \, l'=1}^{2l+1} \sum_{odd \, l''=1}^{l''_{max}} \tilde{l} (\tilde{l} + 1) \int_{r_i}^{r_o}  K_{l'}(r) U_{l''}(r) dr ,
\label{eq:integration_3}
\end{equation}
where $\tilde{l} = \min(l',l'')$.

There is no common natural basis for the radial dependence of the rotational kernels and the azimuthal velocity, and therefore, we use the same equidistant radial grid ($N_r=100$) as in the Tikhonov regularization method.

We are now ready to set up the inversion of the measured mode splittings. Applying the Bayesian formalism recalled in Appendix \ref{Bayesian_formalism}, the data vector $\bm \Delta$ collects the 26 splitting measurements $\Delta_{nlm}$. We assume that the measurement errors are uncorrelated, implying that the data error covariance matrix $C_{dd}$ is diagonal. Its diagonal elements are taken as $\sigma_d^2$ with $\sigma_d =  \epsilon_i + 0.02$ (to account for systematic errors).

The model (or parameter) vector $\boldsymbol{p}$ collects the $U_{l''}$ coefficients at every point of the $r$-grid.
Finally, the $C_{pp}$ matrix elements are chosen as:
\begin{equation}
C_{pp}(i,j) = \frac{\sigma_p^2}{l''^2} exp \left(- \frac{r(i)r(j)}{\delta^2} \right),
\end{equation}
where $\delta$ controls the smoothness of the radial profiles of the $U_{l''}(r)$ coefficients, and $\sigma_p/l''$ controls their amplitude.
Choosing $\sigma_p = 0.02$, $\delta = 0.3$, and $l''_{max} = 9$ (which yields 505 model parameters), and using equation \ref{eq:m_1}, we obtain a model that provides a good fit to the measured mode splittings.

\begin{figure}
\centering
\includegraphics[width=0.8\linewidth]{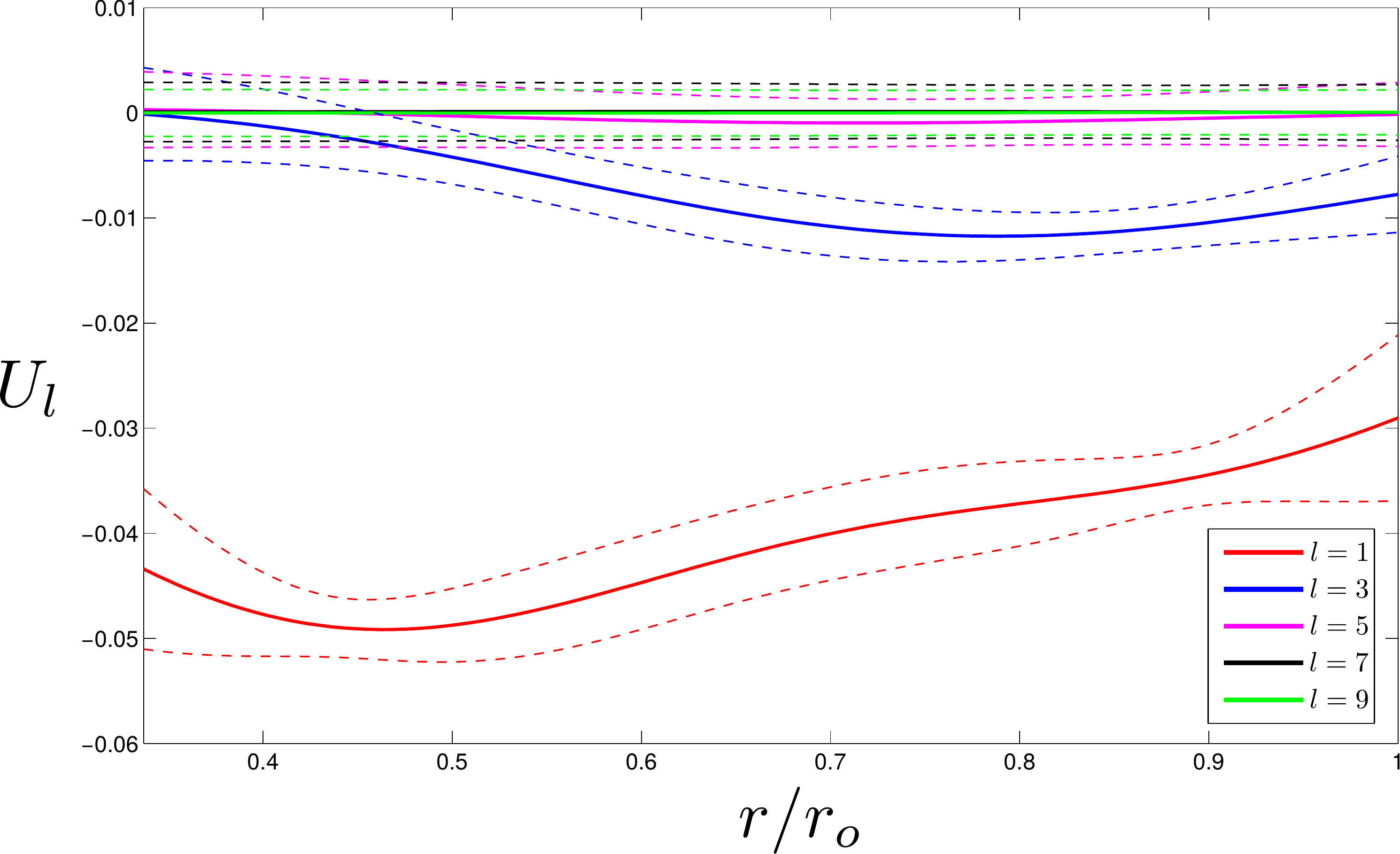}
\caption{Radial profiles of modes $U_{l''}(r)$ of the inverted azimuthal velocity model (semi-spectral inversion). The angular velocity map $\bar \Omega(r,\theta)$ is shown in Fig. \ref{fig:oi} (right quadrant).}
\label{fig:Inverse_10_model}
\end{figure}

Fig. \ref{fig:Inverse_10_model} shows the radial profiles of the $U_{l''}(r)$ coefficients of the inverted model, together with their a posteriori error bar deduced from the a posteriori covariance matrix $C_{\hat{p}\hat{p}}$, computed according to equation \ref{eq:Cpp_1}. Note that our expansion for $U_\varphi (r, \theta)$ does not guarantee that it vanishes on the walls of the shell. This is on purpose since we expect that the boundary conditions $U_\varphi(r_i) = r_i \sin \theta$ and $U_\varphi(r_o) = 0$ are matched through thin viscous boundary layers that cannot be resolved.

The resulting contour map of fluid angular velocity in the $(r, \theta)$ plane is shown on the right quadrant in Fig. \ref{fig:oi}. We note that the model has a dominant $l''=1$ mode (see Fig. \ref{fig:Inverse_10_model}), which corresponds to an angular velocity that does not vary in the $z$-direction. Also note that the angular velocity of the fluid is much lower than that of the spinning inner sphere. Integrating over the fluid volume, we get a dimensionless kinetic energy $E_K \simeq 0.0023$ (solid body rotation of the fluid with the inner sphere would have $E_K = 0.83$ in these $\rho \Omega_i^2 r_o^5$ units).

\begin{table}[p!] 
\caption{\label{theonlytable}Rotationally induced splitting between $\pm m$ modes [mHz/Hz]. }
\begin{ruledtabular}
\begin{tabular}{c c c c}
$(n,l,|m|)$ & Measured $\Delta_{nlm}/\Omega_i$ & Predicted $\bar \Delta_{nlm}/\Omega_i$(Tikhonov) & Predicted $\bar \Delta_{nlm}/\Omega_i$(Bayesian) \\
\hline\\
$(0,1,1)$  & $31   \pm 3$  & $30  $ & $31  \pm 1$ \\
$(0,4,1)$  & $129  \pm 5$  & $110 $ & $110 \pm 8$ \\
$(0,4,4)$  & $281  \pm 3$  & $256 $ & $259 \pm 12$\\
$(0,4,2)$  & $205  \pm 5$  & $204 $ & $200 \pm 9$ \\
$(1,1,1)$  & $155  \pm 3$  & $157 $ & $160 \pm 7$ \\
$(0,5,4)$  & $337  \pm 6$  & $322 $ & $319 \pm 16$\\
$(0,5,2)$  & $216  \pm 16$ & $217 $ & $218 \pm 13$\\
$(0,5,5)$  & $284  \pm 3$  & $299 $ & $303 \pm 19$\\
$(0,5,3)$  & $287  \pm 15$ & $294 $ & $289 \pm 13$\\
$(1,2,1)$  & $202  \pm 2$  & $188 $ & $188 \pm 8$ \\
$(1,2,2)$  & $305  \pm 2$  & $291 $ & $296 \pm 10$\\
$(0,6,3)$  & $293  \pm 13$ & $308 $ & $313 \pm 16$\\
$(1,3,1)$  & $225  \pm 7$  & $187 $ & $88  \pm 10$\\
$(1,3,3)$  & $352  \pm 15$ & $396 $ & $404 \pm 13$\\
$(1,4,1)$  & $183  \pm 5$  & $186 $ & $183 \pm 14$\\
$(2,2,1)$  & $187  \pm 4$  & $195 $ & $193 \pm 9 $\\
$(2,2,2)$  & $268  \pm 18$ & $298 $ & $304 \pm 8 $\\
$(2,3,2)$  & $352  \pm 17$ & $363 $ & $362 \pm 14$\\
$(2,3,3)$  & $494  \pm 3$  & $446 $ & $454 \pm 13$\\
$(1,6,5)$  & $582  \pm 18$ & $585 $ & $594 \pm 29$\\
$(2,4,3)$  & $527  \pm 7$  & $515 $ & $520 \pm 20$\\
$(2,5,5)$  & $705  \pm 34$ & $708 $ & $720 \pm 28$\\
$(3,2,1)$  & $163  \pm 6$  & $188 $ & $185 \pm 9 $\\
$(3,2,2)$  & $261  \pm 6$  & $279 $ & $284 \pm 7 $\\
$(0,13,5)$ & $555  \pm 35$ & $555 $ & $545 \pm 44$\\
$(3,4,4)$  & $568  \pm 25$ & $561 $ & $569 \pm 21$\\
\end{tabular}
\end{ruledtabular}
\end{table}

\begin{figure}
\centering
\includegraphics[width=0.8\linewidth]{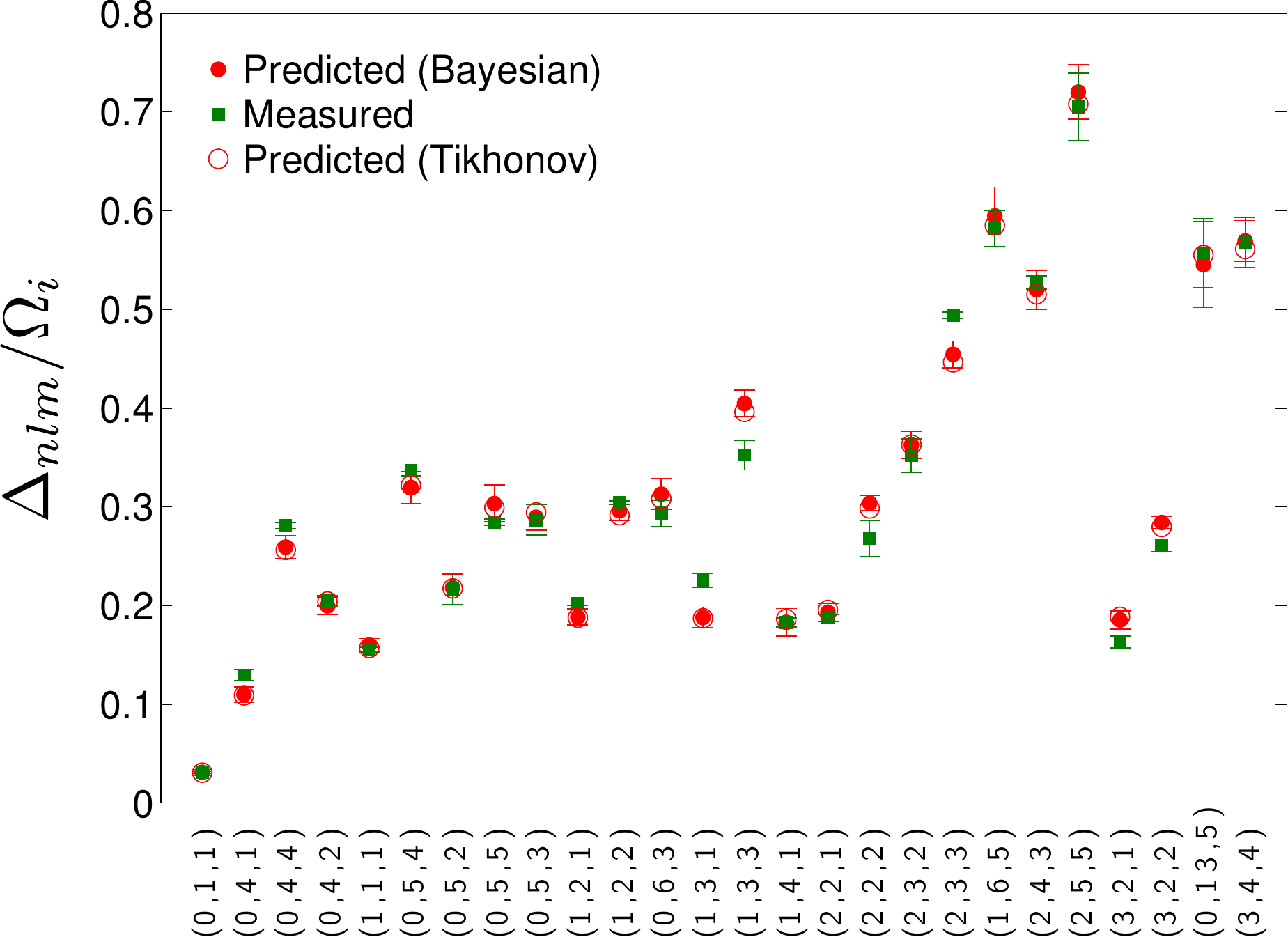}
\caption{Splitting $\Delta_{nlm}/\Omega_i$ of the 26 $(n,l,m)$ modes of section \ref{sec:mi}, with their error bars (green symbols), compared to those predicted by our best-fit models (semi-spectral Bayesian inversion in red filled circles and Tikhonov regularization in red hollow circles).
}
\label{fig:Inverse_10_fits}
\end{figure}

Fig. \ref{fig:Inverse_10_fits} compares the measured mode splittings (with their error bars) to those predicted from our best-fitting model (semi-spectral inversion), with the error bars obtained from the diagonal terms of the $C_{\hat{d}\hat{d}}$ a posteriori covariance matrix.
The overall normalized misfit, as given by equation \ref{eq:misfit}, is $\chi = 0.72$. Table \ref{theonlytable} lists as well the predicted splittings from the best-fitting model along with the predicted splittings obtained from the Tikhonov regularization. 

\section{Concluding remarks}
Techniques from helioseismology are suitable to reconstruct 2-dimensional rotational profiles in experiments involving rotating fluids, as we have demonstrated in this proof-of-concept study. We recovered successfully the angular velocity of the air inside the spherical shell with negligible flow obstruction using low-cost off-the-shelf components. We used two different inversion techniques to reconstruct the azimuthal flow obtaining similar results, except in regions close to the rotation axis where the information provided by the measured splittings was insufficient. The remaining differences on the results can be traced back to the Bayesian inversion's use of norm-damping (via the parameter $\sigma_p$), whereas the Tikhonov's regularization procedure involved only second-order derivative damping.
 
The techniques described in this paper can be applied to rotating flows contained in any axisymmetric container, provided that its acoustic eigenfunctions are known. The eigenfunctions for arbitrary shapes can be readily computed using finite element methods. Another requirement is that a sufficient number of rotational splittings are measurable, which implies that the resonances are narrow enough to be resolved and that the signal to noise ratio is adequate. Ideally one would want to measure as many splitting modes as possible since then the averaging kernels would be more localized, improving spatial resolution and reducing the error magnification.

One can envisage a scheme to force the set of all measurable splitting eigenmodes via targeted, short and simultaneous chirps, in such a way that an inversion can be computed in a timely fashion, thus allowing some temporal monitoring of the flow. In a forthcoming study we explore the possibility of using accelerometers instead of microphones, in which case the eigenfunctions correspond to the combined acoustic forcing and elastic response of the bounding cavity. The use of accelerometers is convenient when the inner surface of the bounding cavity is not accessible or when the microphones are not compatible with the fluid.

This study was originally motivated by the need of a quantitative measurement of the internal differential rotation in the 3-meter diameter, liquid-metal experiment at the University of Maryland. Inertial waves can be easily excited by differential rotation in this device \citep{triana2011inertial} and a measure of the rotational profile using helioseismic techniques will allow a comparison with recent theoretical \citep{rieutord2012excitation} and numerical studies \citep{wicht2014flow}, thus gaining further insight into the inertial mode excitation mechanism. The use of this technique will also improve the physical picture behind the bi-stability phenomenon observed in the 3-meter device \citep{zimmerman2011bi}. It is worth noting, from the theoretical astrophysics viewpoint, that both inertial wave excitation and the bi-stability phenomenon are intimately related to angular momentum transport processes within the fluid, a topic which little is known but is of fundamental importance in solar and stellar physics.       

\appendix

\section{\label{ap:acmo}Acoustic eigenmodes of a spherical shell}

The spatial component $f(\bm r)$ of an acoustic eigenmode obeys the Helmholtz equation
$$
\nabla^2 f(\bm r)+k^2 f(\bm r)=0,
$$
where $k=\omega/c$ is the wave number. This equation is separable in spherical coordinates so that $f(\bm r)=R(r)Q(\theta)\Phi(\phi)$. The solutions for the equations on $\theta$ and $\phi$ are the familiar spherical harmonics $Y_l^m(\theta,\phi)$ and the radial solution $R(r)$ is \cite{russell2010basketballs}, apart from a normalizing factor,
$$
R(r)=j_l(k r)+B n_l(k r),
$$
where $j_l$ and $n_l$ are the spherical Bessel functions (order $l$) of the first and second kind, respectively. The constant $B$ and the wave number $k$ are determined by the boundary conditions $R'(r=r_i)=0$ and $R'(r=r_o)=0$ (primes denote derivatives). For each $l$ there is a numerable infinite collection of $(k,B)$ pairs allowed by the boundary conditions. To distinguish among them we arrange them in ascending order according to the magnitude of $k$ and label them with the index $n$ starting from $n=0$. So we can write
$$
f_{nlm}(r,\theta,\phi)=\left[j_l(k_{nl} r)+B_{nl}(k_{nl} r)\right]Y_l^m(\theta,\phi).
$$  
Note that the azimuthal wave number $m$ does not enter in the radial equation defining $k$, therefore a given $(n,l)$ mode is $2l+1$ degenerate.

Now we want the corresponding functional form of the displacement of a fluid particle $\bm \zeta (\bm r)$, a necessary quantity to compute the kernels. $\bm \zeta$ is related to the velocity $\bm u$ through $\bm u = \partial \bm \zeta/\partial t$. Assuming again an harmonic time dependence ${\rm} e^{{\rm -i} \omega t}$, the Euler equation leads to
$$
\bm \zeta = \frac{\nabla P}{\rho \omega^2}.
$$
After replacing the pressure $P$ by the pressure eigenfunction $P_{nlm}$ one gets
$$
\bm \zeta = \zeta_r \hat{\bm r} + \bm \zeta_h,
$$
where the radial displacement $\zeta_r$ is given by
$$
\zeta_r = \frac{1}{\rho \omega^2}R'(r) P_l^m(\cos \theta) e^{i m \phi},
$$
and the horizontal displacement is given by
$$
\bm \zeta_h = \frac{1}{\rho \omega^2}\frac{R(r)}{r}\left[\frac{d}{d \theta}P_l^m(\cos \theta) \hat{\bm \theta} + i\frac{m}{\sin \theta} P_l^m(\cos \theta) \hat{\bm \phi} \right]e^{i m \phi},
$$
The functions $\xi_r$ and $\xi_h$ involved in the kernel calculation (see next appendix) are given by
$\xi_r=R'(r)$ and $\xi_h=R(r)/r$.

\section{\label{ap:kern}Kernels and splittings}

A detailed derivation of the rotational splitting can be found in \cite{aerts2010asteroseismology} (section 3.8.3, page 271). Here we list the main formulae.
If the rotation frequency of the fluid $\Omega/2 \pi$ is small compared to the mode's eigenfrequency then the frequency splitting is given by
$$
\Delta_{nlm} = m \frac{R_{nlm}}{I_{nlm}},
$$
which can be written equivalently as
$$
\Delta_{nlm}= m \int_{r_i}^{r_o} \int_0^\pi K_{nlm}(r,\theta)\, \Omega(r,\theta)\, r dr d\theta.
$$
$R_{nlm}$ and $I_{nlm}$ are calculated from $\xi_r$ and $\xi_h$ (see previous appendix) as follows:
\begin{widetext}
\begin{equation}
\begin{split}
R_{nlm} = \int_{r_i}^{r_o} \int_0^\pi \left\{ \vphantom{\frac{a^2}{b^2}} \right. & |\xi_r(r)|^2 p^2 + |\xi_h(r)|^2 \left[q+\frac{m^2}{\sin^2\theta}p^2\right]+ \\
& -p^2 \left[\xi_r^*(r) \xi_h(r)+\xi_r(r) \xi_h^*(r)\right] \\ 
& -2 pq \frac{|\xi_h(r)|^2}{\tan \theta}
\left. \vphantom{\frac{a^2}{b^2}}\right\} \Omega(r,\theta)\, \rho(r)\, r^2 \sin \theta \, d\theta \,dr 
\end{split}
\end{equation}
\begin{equation}
I_{nlm} = \int_{r_i}^{r_o} \int_0^\pi \left\{ |\xi_r(r)|^2 p^2 + |\xi_h(r)|^2 \left[q+\frac{m^2}{\sin^2\theta}p^2\right] \right\}  \rho(r)\, r^2 \sin \theta \, d\theta \,dr, 
\end{equation}
\end{widetext}
where we have defined $p\equiv P_l^m(\cos \theta)$ and $q\equiv \frac{d}{d \theta}P_l^m(\cos \theta)$. Now if we use the fully normalized associated Legendre polynomials, i.e. those for which
$$\int_0^\pi P_l^m(\cos \theta)^2 \sin \theta \, d\theta=1,$$
we can express $I_{nlm}$ more compactly as 
$$
I_{nlm}=\int_{r_i}^{r_o} \left[ |\xi_r(r)|^2  + l(l+1)|\xi_h(r)|^2 \right] \, r^2 \,dr,
$$
assuming $\rho(r)=1$. Finally, we write the rotational kernel as

\begin{widetext}
	\begin{equation}
		K_{nlm}(r,\theta)=\frac{r \sin \theta}{I_{nlm}} \left\{\xi_r^2 p^2+\xi_h^2 \left[q^2+\frac{m^2}{\sin^2 \theta}p^2 -2\frac{p q}{\tan \theta} \right] -2\xi_h \xi_r p^2 \right\}.
	\end{equation}
\end{widetext}

We can take advantage of the axisymmetry of the kernels. They can be projected on $P_{l'}^1(\cos \theta)$ associated Legendre polynomials of degree $l'$ and order $1$. Since they involve quadratic terms of the pressure field, which was expanded on $Y_{l}^m(\theta, \phi)$, and because they must have $m \ge 1$, the expansion is exact when extended to degree $l'_{max} = 2 l+1$.

The kernels are symmetrical with respect to the equator, implying that only odd degrees intervene in this expansion. We thus have:

\begin{equation}
K_{nlm}(r,\theta) = \sum_{odd \; l'=1}^{2l+1} K_{l'}(r) P_{l'}^1(\cos \theta),
\label{eq:Legendre_K}
\end{equation}

This expansion of $K_{nlm}(r,\theta)$ in $P_{l'}^1(\cos \theta)$ is used in the semi-spectral inversion for the velocity field, taking advantage of the fact that the $P_{l}^1(\cos \theta)$ are also the natural basis for the average azimuthal velocity field, which is axisymmetric and symmetric with respect to the equator as well.

\section{\label{ap:tr}Tikhonov regularization}

Consider a $(r,\theta)$ grid with $N_r+1$ points in the radial direction and $N_\theta+1$ points in the $\hat \theta$ direction. The goal is to determine the $N_r \times N_\theta$ unknowns $\bar \Omega_{\alpha \beta}$ which represent the predicted angular velocity $\bar \Omega$ at radius $r$ such that $r_{\alpha-1}\leq r \leq r_\alpha$ and colatitude $\theta$ such that $\theta_{\beta-1}\leq \theta \leq \theta_\beta$, where $\alpha=1,\ldots,N_r$ and $\beta=1,\ldots,N_\theta$ are the grid indices. The Tikhonov regularization method seeks to minimize the quantity $\mathscr{T}$ defined in Eq. \ref{eq:Tikh0}, through
\begin{equation}
\frac{\partial \mathscr{T}}{\partial \bar \Omega_{\alpha \beta}}=0.
\label{eq:Tikh2}
\end{equation}
By using the discretization of $\bar \Omega$ over the $(r,\theta)$ grid, the integrals become discrete sums. In particular, the predicted splitting $\bar \Delta_i$ becomes
\begin{equation}
	\bar \Delta_i=2 m_i \sum_{\alpha,\beta}\int\limits_{r_{\alpha-1}}^{r_\alpha} \int\limits_{\theta_{\beta-1}}^{\theta_\beta} K_i(r,\theta)\, \bar\Omega_{\alpha \beta}\, r\, \mathrm{d}r\, \mathrm{d}\theta.
\label{eq:deltai}
\end{equation}
Note that we need to consider a grid only on a quadrant $(0\leq \theta \leq \pi/2)$ since the kernels $K_i(r,\theta)$ are symmetric with respect to the equator and we assume $\Omega(r,\theta)$ sharing this symmetry as well. Therefore the factor $2$ on the right hand side of Eq. \ref{eq:deltai}.
If we 'unzip' the unknowns $\bar \Omega_{\alpha,\beta}$ to form a single column vector with $N_r \times N_\theta$ elements total and running index $j$, we can then rewrite Eq. \ref{eq:deltai} as
\begin{equation}
    \bar \Delta_i=\sum_{j=1}^{N_r\times N_\theta} G_{i j}\, \bar \Omega_j,
\end{equation}
where the matrix element $G_{i j}$ is
\begin{equation}
G_{i j}=2 m_i \int\limits_{r_{\alpha(j)-1}}^{r_{\alpha(j)}} \int\limits_{\theta_{\beta(j)-1}}^{\theta_{\beta(j)}} K_i(r,\theta)\, r\, \mathrm{d}r\, \mathrm{d}\theta \qquad j=1,\ldots, N_r \times N_\theta \qquad i=1,\ldots,M.    
\end{equation}
Similarly, we construct the discrete versions of the last two terms of Eq. \ref{eq:Tikh0}. After some careful manipulation (necessary given the 'unzipping' of the grid) we can write the matrix equivalent of Eq. \ref{eq:Tikh2} as
\begin{equation}
\mathbf{Q}^\top \bm \Delta - \left( \mathbf{Q}^\top\mathbf{G} +\mu_r\frac{\delta \theta}{\delta r^3}\mathbf{L}_r+\mu_\theta \frac{\delta r}{\delta\theta^3}\mathbf{L}_\theta \right) {\bar{\bm \Omega}} = 0
\label{eq:Tikh_mat}
\end{equation}
where $\mathbf{G}$ is a $M\times(N_r \times N_\theta)$ matrix, $\mathbf{Q}$ is such that $Q_{ij}=G_{ij}/\epsilon_i^2 $, $\delta r$ and $\delta \theta$ are the grid spacings, $\mathbf{L}_r$ and $\mathbf{L}_\theta$ are the matrices related to the discrete derivatives along $r$ and $\theta$, $\bm \Delta$ is the vector of all the measured splittings ($M$ in total), and $\bar{\bm \Omega}$ is the column vector of $N_r \times N_\theta$ unknowns.
Additionally, we require $\bar \Omega$ constrained so that
\begin{equation} 
\left. \frac{\partial \bar \Omega}{\partial \theta}\right|_{\theta=\pi/2}=0,
\label{eq:lagr}
\end{equation}
ensuring a smooth and symmetric solution across $\theta=\pi/2$. This is accomplished by introducing $N_r$ Lagrange multipliers (appended to $\bar{\bm \Omega}$, resizing all matrices accordingly) together with the matrix version of Eq. \ref{eq:lagr} added to Eq. \ref{eq:Tikh_mat}.

The final step involves the calculation of the inverse matrix (in the least squares sense)
\begin{equation}
\mathbf{T}^{-1}=\left( \mathbf{Q}^\top\mathbf{G} +\mu_r\frac{\delta \theta}{\delta r^3}\mathbf{L}_r+\mu_\theta \frac{\delta r}{\delta\theta^3}\mathbf{L}_\theta + \mathbf{H}\right)^{-1},
\end{equation}
where $\mathbf{H}$ is the matrix representing the constraint in Eq. \ref{eq:lagr}. Finally we can write the solution for the inferred angular velocity as
\begin{equation}
\bar{\bm \Omega}=\mathbf{T}^{-1}\mathbf{Q}^\top \bm \Delta
\label{eq:TO}
\end{equation}

Let us now examine how the selection of the trade-off parameters $\mu_r$ and $\mu_\theta$ affect the inversion. Including the experimental error $\epsilon_i$ in measuring the splitting $\Delta_i$ we write:
\begin{equation}
\Delta_i= m \int_{r,\theta} K_i(r,\theta)\, \Omega(r,\theta)\, r\, \mathrm{d}r\, \mathrm{d}\theta + \epsilon_i,
\end{equation}
where $\Omega(r,\theta)$ is the true angular velocity. An approximation $\bar \Omega$ to the true angular velocity at a given point $(r',\theta')$ is linearly related to the splitting data (Eq. \ref{eq:TO}). At each $(r',\theta')$ we can find a set of {\it inversion coefficients} $c_i(r',\theta')$ satisfying:
\begin{equation}
\bar \Omega(r',\theta')=\sum_i c_i(r',\theta') \Delta_i=\int_{r,\theta}\mathcal{K}(r',\theta',r,\theta)\, \Omega(r,\theta)\, r\, \mathrm{d}r\, \mathrm{d}\theta+\sum_i c_i(r',\theta')\, \epsilon_i.
\label{eq:inv_coeff}
\end{equation}
The {\it averaging kernel} $\mathcal{K}$ is then related to the rotational kernels $K_i$ through
\begin{equation}
\mathcal{K}(r',\theta',r,\theta)=\sum_i c_i(r',\theta')\, K_i(r,\theta).
\label{eq:avker}
\end{equation}
Ideally, the averaging kernel $\mathcal{K}$ is sharply peaked around $(r',\theta')$ with small values away from that point. The more modes are considered, the better the localization of $\mathcal{K}$ becomes. The resolution of the inversion at a given point $(r',\theta')$ can be characterized by the width of the averaging kernel at that point, both in the $\hat r$ and $\hat \theta$ direction. The widths are usually expressed as the difference between the first and third quartile points. In the radial direction this means the quantity $\rho_3-\rho_1$, where $\rho_k$ is defined by
\begin{equation}
\int_{r_i}^{\rho_k}\mathcal{K}(r',\theta',r,\theta')\,\mathrm{d}r=\frac{k}{4}\int_{r_i}^{r_o}\mathcal{K}(r',\theta',r,\theta')\,\mathrm{d}r,
\label{eq:width}
\end{equation}
and similarly for the angular width $\theta_3-\theta_1$. The variance in the result of the inversion at $(r',\theta')$, assuming uncorrelated standard errors $\epsilon_i$ on $\Delta_i$ is
\begin{equation}
\sigma^2 \left[\bar \Omega(r',\theta')\right]=\sum_i c_i^2(r',\theta')\, \epsilon_i^2,
\label{eq:acc}
\end{equation}
In our case the errors are more or less the same for all modes, $\epsilon_i=\epsilon$, so we can express the standard deviation as $\sigma \left[\bar \Omega(r',\theta')\right]=\Lambda(r',\theta')\, \epsilon$, where the {\it error magnification} $\Lambda(r',\theta')$ is defined as
\begin{equation}
\Lambda(r',\theta')=\left[\sum_i c_i^2(r',\theta')\right]^{1/2}.
\label{eq:ermag}
\end{equation}
Generally speaking, small trade-off parameters lead to high error magnifications (over-fitting) and small {\it rms} error between observed and predicted splittings. Increasing the value of the trade-off parameters would improve the error magnification at the cost of decreased spatial resolution, i.e., larger averaging kernel widths.

\section{\label{Bayesian_formalism}Bayesian formalism}

Within our current approximation, the splitting is linearly related to the fluid velocity and we can use the classical formalism of generalized least-square linear inversion. Following \cite{tarantola1982generalized} we define a model vector $\boldsymbol{m}$ and a data vector $\bm \Delta$. The data vector $\bm \Delta$ collects our splitting measurements. The fluid velocity field is projected on a semi-spectral basis, and the model vector $\boldsymbol{m}$ collects the coefficients of the model solution for this projection. We assume that the measurement errors obey a gaussian statistics, and define $C_{dd}$ the covariance matrix of the data. A priori information on the model parameters is given through an a priori covariance matrix $C_{pp}$ on the model parameters. It will be used to control the smoothness of the desired model. Finally, the $\mathbf{G}$ matrix defines the linear relationship between the model parameters vector $\boldsymbol{m}$  and the data vector $\bm \Delta$, so that:
\begin{equation}
\bm \Delta = \mathbf{G} \boldsymbol{m}. 
\end{equation}
Within this Bayesian framework, the best fitting model $\boldsymbol{\hat{m}}$, in a least-square sense, is given by \citep{tarantola1982generalized}:
\begin{equation}
\boldsymbol{\hat{m}} = C_{pp} \mathbf{G}^T \left(C_{dd} + \mathbf{G} C_{pp} \mathbf{G}^T \right)^{-1} \bm \Delta,
\label{eq:m_1}
\end{equation}
or equivalently by:
\begin{equation}
\boldsymbol{\hat{m}} = \left(\mathbf{G}^T C_{dd}^{-1} \mathbf{G} + C_{pp}^{-1} \right)^{-1} \mathbf{G}^T C_{dd}^{-1} \bm \Delta.
\label{eq:m_2}
\end{equation}
The accuracy of this best fitting model is assessed through the a posteriori covariance matrix of the parameters, given by:
\begin{equation}
C_{\hat{p}\hat{p}} = C_{pp}  - C_{pp} \mathbf{G}^T \left(C_{dd} + \mathbf{G} C_{pp} \mathbf{G}^T \right)^{-1} \mathbf{G} C_{pp},
\label{eq:Cpp_1}
\end{equation}
or equivalently by:
\begin{equation}
C_{\hat{p}\hat{p}} = \left(\mathbf{G}^T C_{dd}^{-1} \mathbf{G} + C_{pp}^{-1} \right)^{-1}.
\label{eq:Cpp_2}
\end{equation}

The diagonal elements of this matrix provide the variance of the model parameters. However, the a posteriori covariance matrix is not diagonal, and it is usual to analyse the a posteriori information via a resolution matrix $R$, which indicates how well the data resolve the parameters of the model. As noted by \cite{jackson1979the}, this can be expressed as $\boldsymbol{\hat{m}} = R \boldsymbol{m}$, where $\boldsymbol{m}$ is the `true' model such that $\bm \Delta = \mathbf{G} \boldsymbol{m}$. Equations \ref{eq:m_1} and \ref{eq:m_2} thus yield two possible expressions for $R$:
\begin{equation}
R = C_{pp} \mathbf{G}^T \left(C_{dd} + \mathbf{G} C_{pp} G^T \right)^{-1} \mathbf{G} = \left(\mathbf{G}^T C_{dd}^{-1} \mathbf{G} + C_{pp}^{-1} \right)^{-1} \mathbf{G}^T C_{dd}^{-1} \mathbf{G}.
\label{eq:resolution}
\end{equation}

When examining the $i^{th}$ parameter of our best-fitting model $\boldsymbol{\hat{m}}$, we will keep in mind that the data is only providing a blurred view of this parameter, through a filter given by the $i^{th}$-line of the resolution matrix $R$.

We define the mean normalized data misfit $\chi$ as:
\begin{equation}
\chi = \frac{1}{M} \sqrt{(\bm \Delta - \mathbf{G} \boldsymbol{\hat{m}})^T C_{dd}^{-1} (\bm \Delta - \mathbf{G} \boldsymbol{\hat{m}})},
\label{eq:misfit}
\end{equation}
where $M$ is the number of splitting measurements.

The splittings predicted by the best-fitting model are given by $\bm {\hat{\Delta}} = \mathbf{G} \boldsymbol{\hat{m}}$, with an a posteriori covariance matrix $C_{\hat{d} \hat{d}} = \mathbf{G} C_{\hat{p} \hat{p}} \mathbf{G}^T$.

\begin{acknowledgments}
We thank Conny Aerts for valuable comments. S.A. Triana gladly acknowledges support through a Pegasus Marie Curie Fellowship from the Research Foundation - Flanders (FWO). The Grenoble team acknowledges support from LabEx OSUG@2020 (Investissements d'avenir ANR10 LABX56) and thanks A. Richard and Ph. Roux for their help. This collaborative work was initiated thanks to support by the University of Maryland, CNRS and University of Grenoble under Memorandum of Understanding CNRS-723684/00.
\end{acknowledgments}

\bibliography{acoustic_paper2}

\end{document}